\documentclass[a4paper, 11pt]{article}
\usepackage[left=2.5cm,right=2.5cm,top=2.5cm,bottom=2.5cm,includefoot]{geometry}

\usepackage[british,UKenglish,USenglish,english,american]{babel}

\usepackage{dsfont} % contains the command \mathds{1} for the indicator function
\usepackage[font={footnotesize}]{caption}
\usepackage[font={footnotesize}]{subcaption} % for subfigures
\usepackage{graphicx}
\usepackage{booktabs}
\usepackage{amsmath}
\usepackage{amssymb}
\usepackage{bbold}
\usepackage{booktabs}
\usepackage{lineno}
\usepackage{multicol}
\usepackage{multirow}
\usepackage{authblk}
\usepackage[section]{placeins}
\usepackage{lipsum}

\usepackage{url}
\makeatletter
\g@addto@macro{\UrlBreaks}{\UrlOrds}
\makeatother

\usepackage[colorlinks=true, linkcolor=blue, citecolor=blue, urlcolor=blue, linktocpage=true]{hyperref} 

\usepackage{mathtools}

\newtagform{brackets}{[}{]}
\usetagform{brackets}

\newcommand{\one}{\mathds{1}}
\newcommand{\real}{\mathbb{R}}

\newcommand{\myS}{\textsf{S}}
\newcommand{\SX}{{\bar{\myS}_{\textsf{X}}}}
\newcommand{\SC}{{\bar{\myS}_{\textsf{C}}}}
\newcommand{\SR}{{\bar{\myS}_{\textsf{R}}}}

\newcommand{\MCB}{\textsf{MCB}}
\newcommand{\DSC}{\textsf{DSC}}
\newcommand{\UNC}{\textsf{UNC}}

\newcommand{\REL}{\textsf{REL}}
\newcommand{\RES}{\textsf{RES}}

\newcommand*{\QEDB}{\null\nobreak\hfill\ensuremath{\square}}%

\newcommand\blfootnote[1]{%
	\begingroup
	\renewcommand\thefootnote{}\footnote{#1}%
	\addtocounter{footnote}{-1}%
	\endgroup
}

\title{Evaluating probabilistic classifiers: Reliability diagrams and score decompositions revisited}

\author[1,2]{Timo Dimitriadis}
\author[1,3]{Tilmann Gneiting}
\author[1]{Alexander I. Jordan}

\affil[1]{Computational Statistics (CST) Group, Heidelberg Institute for Theoretical Studies, Heidelberg, Germany}
\affil[2]{Institute of Economics, University of Hohenheim, Stuttgart, Germany}
\affil[3]{Institute for Stochastics, Karlsruhe Institute of Technology (KIT), Karlsruhe, Germany}
\begin{document}
	
	\maketitle

	\begin{abstract}
		A probability forecast or probabilistic classifier is reliable or
		calibrated if the predicted probabilities are matched by ex post
		observed frequencies, as examined visually in reliability diagrams.
		The classical binning and counting approach to plotting reliability
		diagrams has been hampered by a lack of stability under unavoidable,
		ad hoc implementation decisions.  Here we introduce the CORP approach,
		which generates provably statistically Consistent, Optimally binned,
		and Reproducible reliability diagrams in an automated way.  CORP is
		based on non-parametric isotonic regression and implemented via the
		Pool-adjacent-violators (PAV) algorithm --- essentially, the CORP
		reliability diagram shows the graph of the PAV-(re)calibrated forecast
		probabilities.  The CORP approach allows for uncertainty
		quantification via either resampling techniques or asymptotic theory,
		furnishes a new numerical measure of miscalibration, and provides a
		CORP based Brier score decomposition that generalizes to any 
		proper scoring rule.  We anticipate that judicious uses of the PAV
		algorithm yield improved tools for diagnostics and inference for a
		very wide range of statistical and machine learning methods.
	\end{abstract}

	\blfootnote{We thank Andreas Fink, Peter Knippertz, Peter Vogel and
		seminar participants at MMMS2 for providing data, discussion and
		encouragement.  Our work has been supported by the Klaus Tschira
		Foundation, by the University of Hohenheim, by the Helmholtz
		Association, and by the Deutsche Forschungs\-ge\-mein\-schaft (DFG,
		German Research Foundation) -- Project-ID 257899354 -- TRR 165.}

 \newpage
 	
 \section{Introduction} 
	
	{C}alibration or reliability is a key requirement on any
	probability forecast or probabilistic classifier.  In a nutshell, a
	probabilistic classifier assigns a predictive probability to a binary
	event.  The classifier is calibrated or reliable if, when looking back
	at a series of extant forecasts, the conditional event frequencies
	match the predictive probabilities.  For example, if we consider all
	cases with a predictive probability of about .80, the observed event
	frequency ought to be about .80 as well.  While for many decades
	researchers and practitioners have been checking calibration in
	myriads of applications (\ref{Spiegelhalter1986}, \ref{Murphy1992}),
	the topic is subject to a surge of interest in machine learning
	(\ref{Flach2016}), spurred by the recent recognition that ``modern
	neural networks are uncalibrated, unlike those from a decade ago''
	(\ref{Guo2017}).
	
\section{Reliability diagrams: Binning and counting} 

The key diagnostic tool for checking calibration is the reliability
diagram, which plots the observed event frequency against the
predictive probability.  In discrete settings where there are only a
few predictive probabilities, such as, e.g., $0, \frac{1}{10}, \ldots,
\frac{9}{10}, 1$, this is straightforward.  However, statistical and
machine learning approaches to binary classification generate
continuous predictive probabilities that can take any value between 0
and 1, and typically the forecast values are pairwise distinct.  In
this ubiquitous setting, researchers have been using the ``binning and
counting'' approach, which starts by selecting a certain, typically
arbitrary number of bins for the forecast values.  Then, for each bin,
one plots the respective conditional event frequency versus the
midpoint or average forecast value in the bin.  For calibrated or
reliable forecasts the two quantities ought to match, and so the
points plotted ought to lie on, or close to, the diagonal
(\ref{Murphy1992}, \ref{Broecker2008}).

\begin{figure}[p]
	\centering
	\begin{subfigure}{.4\linewidth}
		\includegraphics[width=\linewidth]{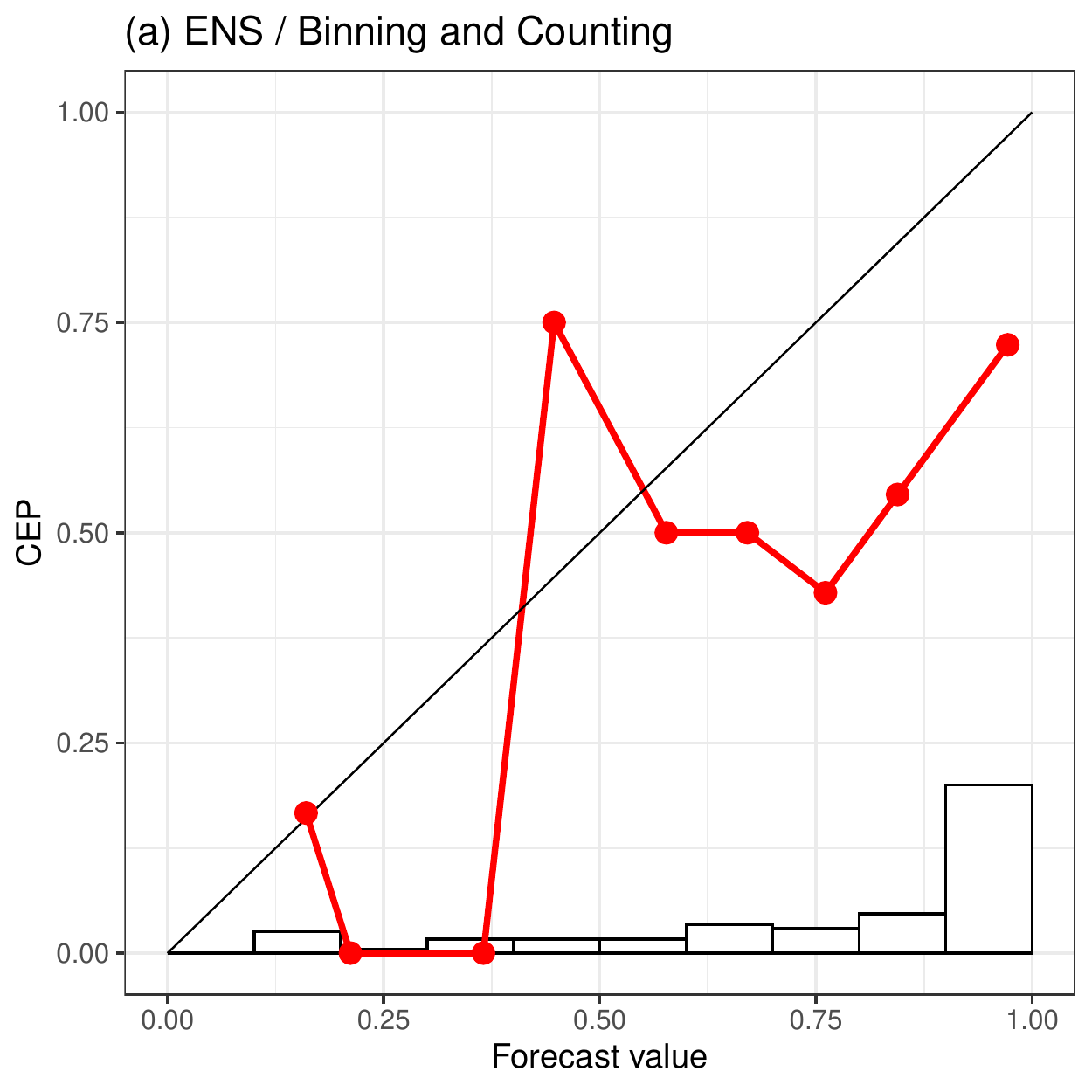} 
	\end{subfigure}
	\begin{subfigure}{.4\linewidth}
		\includegraphics[width=\linewidth]{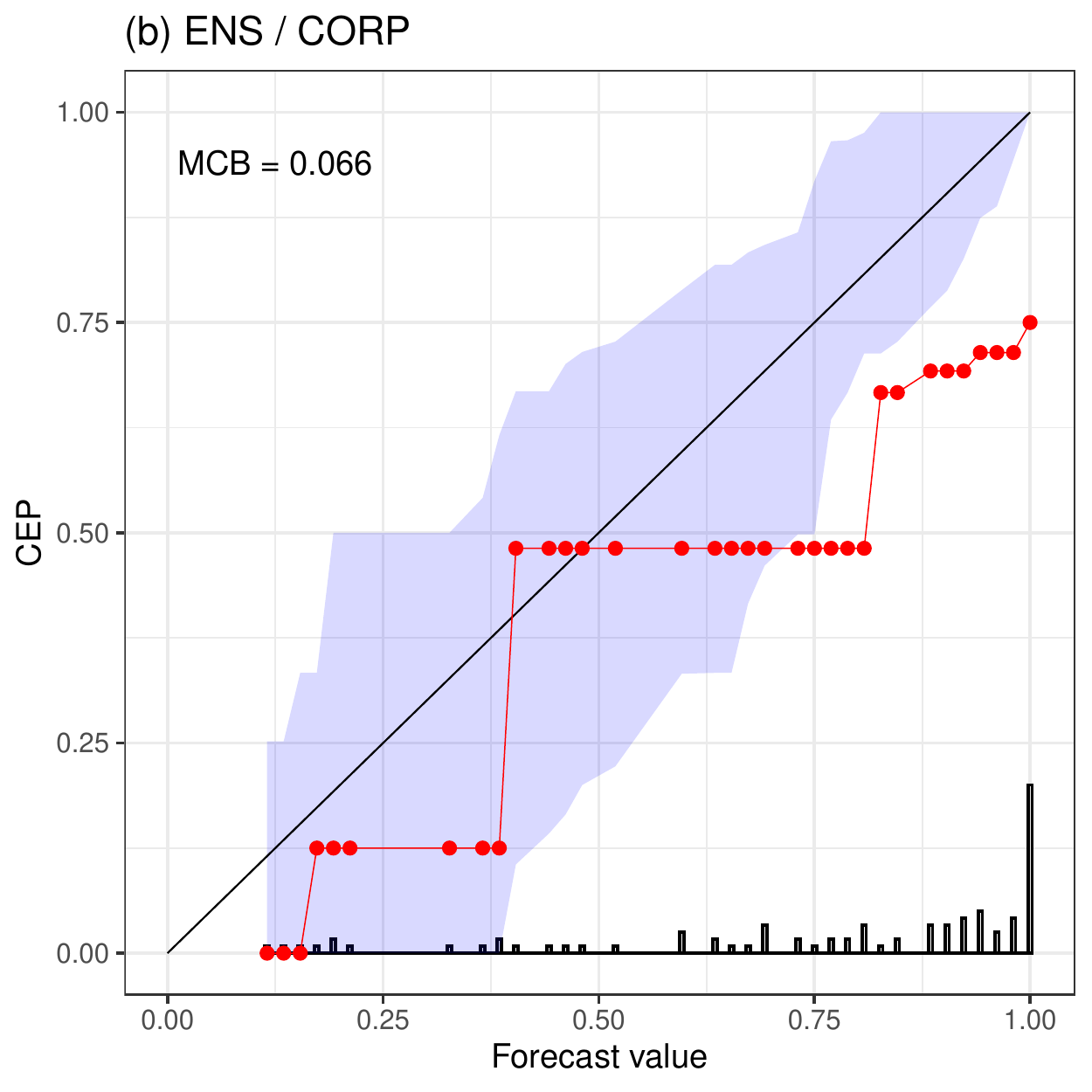} 
	\end{subfigure}
	\begin{subfigure}{.4\linewidth}
		\includegraphics[width=\linewidth]{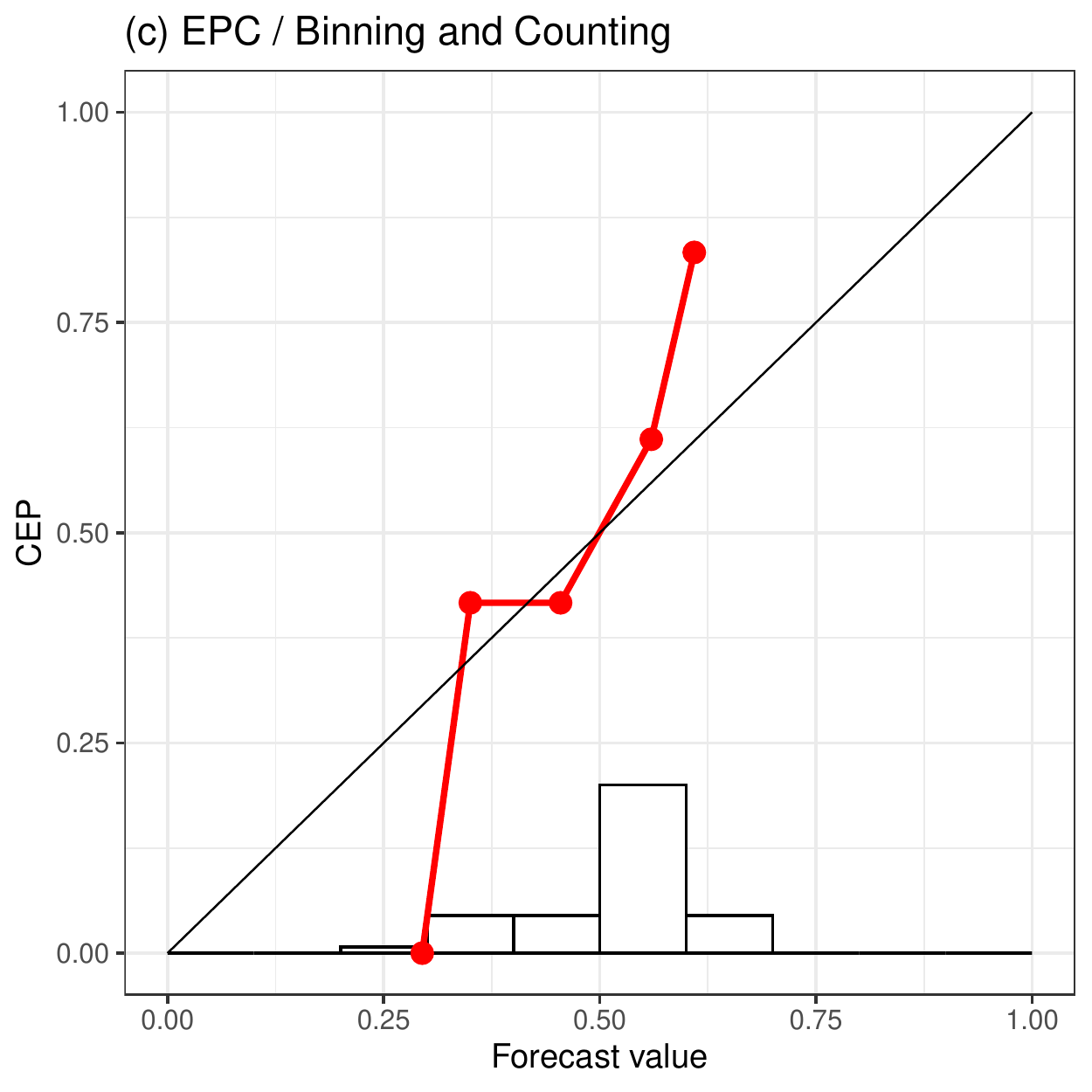}
	\end{subfigure}
	\begin{subfigure}{.4\linewidth}
		\includegraphics[width=\linewidth]{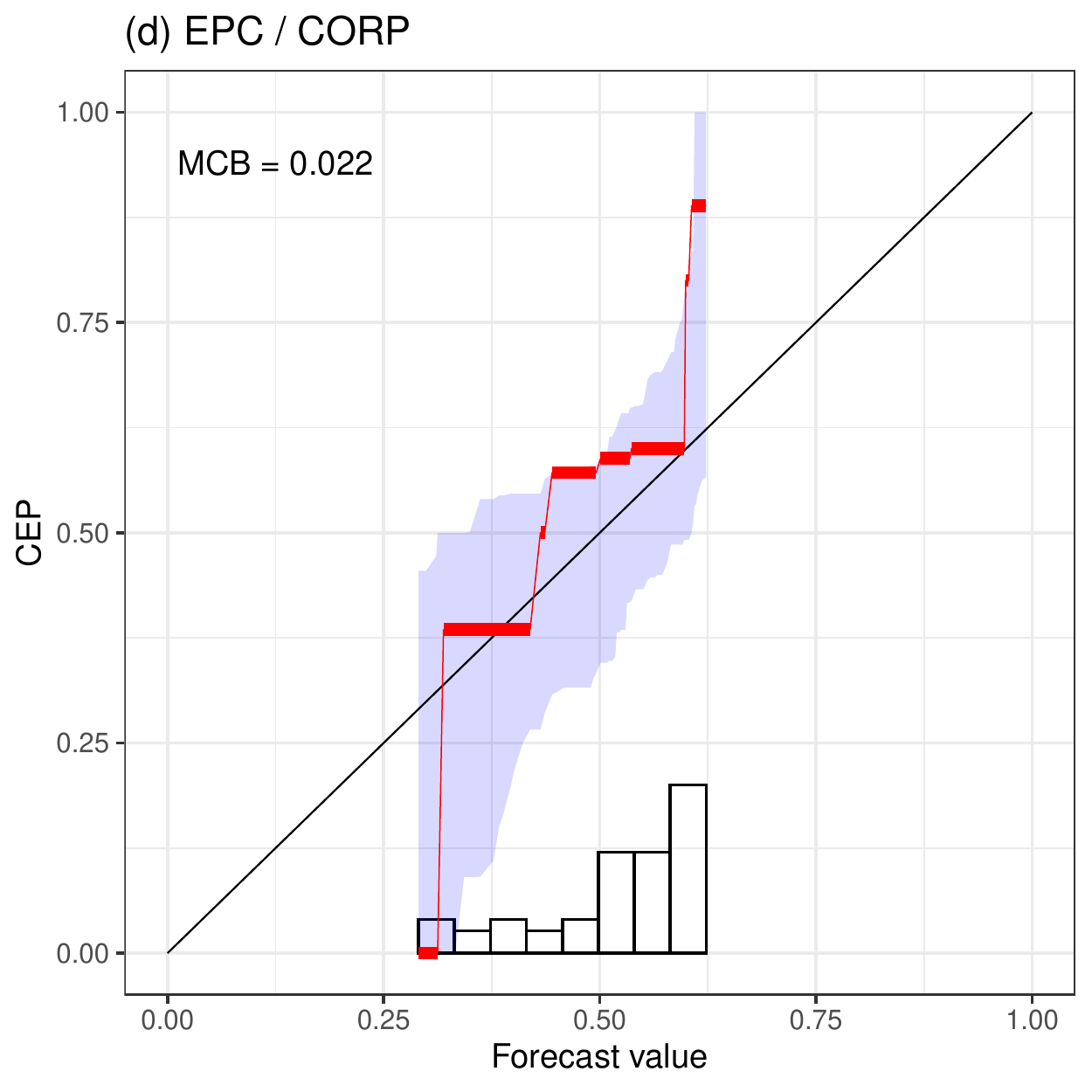}
	\end{subfigure}
	\begin{subfigure}{.4\linewidth}
		\includegraphics[width=\linewidth]{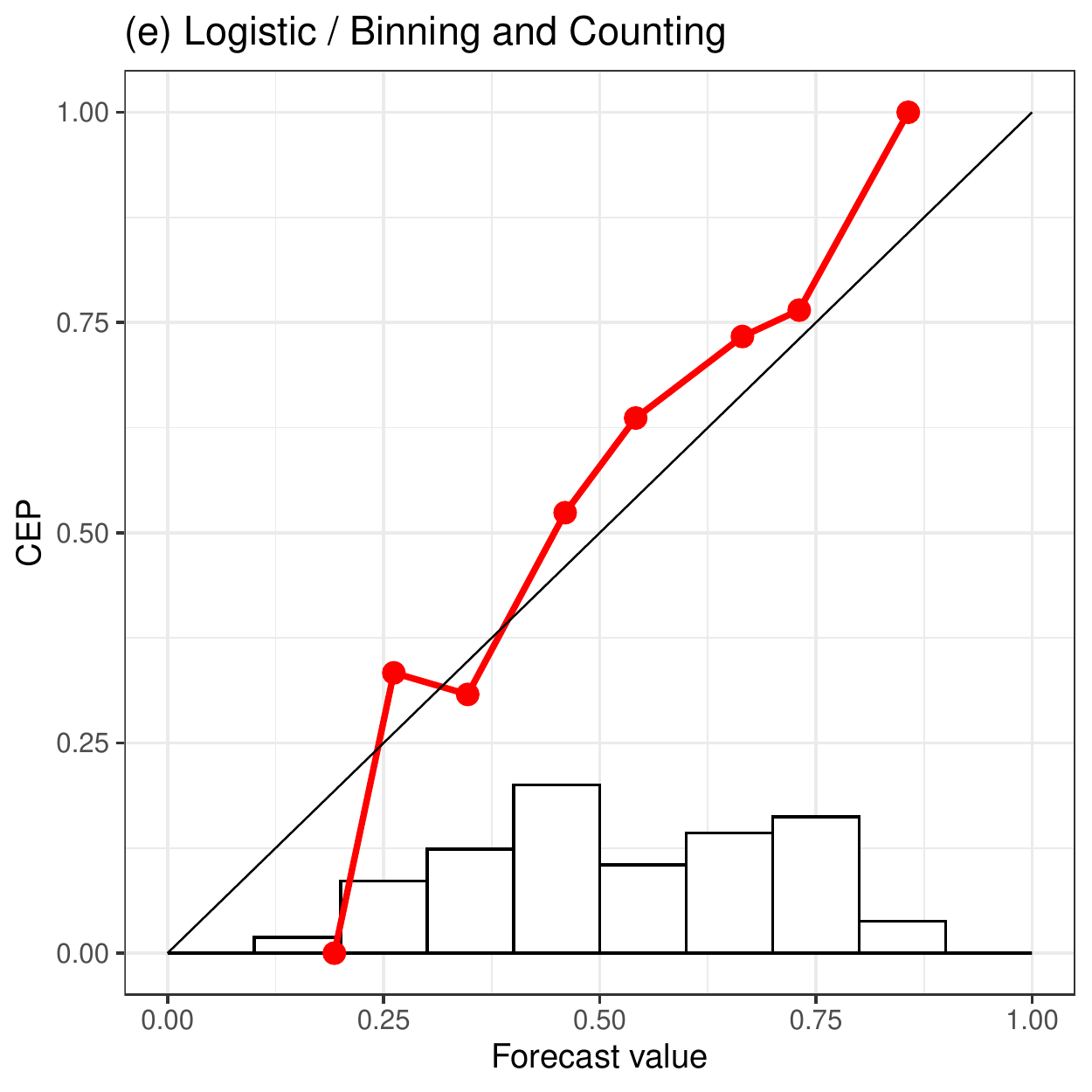}
	\end{subfigure}
	\begin{subfigure}{.4\linewidth}
		\includegraphics[width=\linewidth]{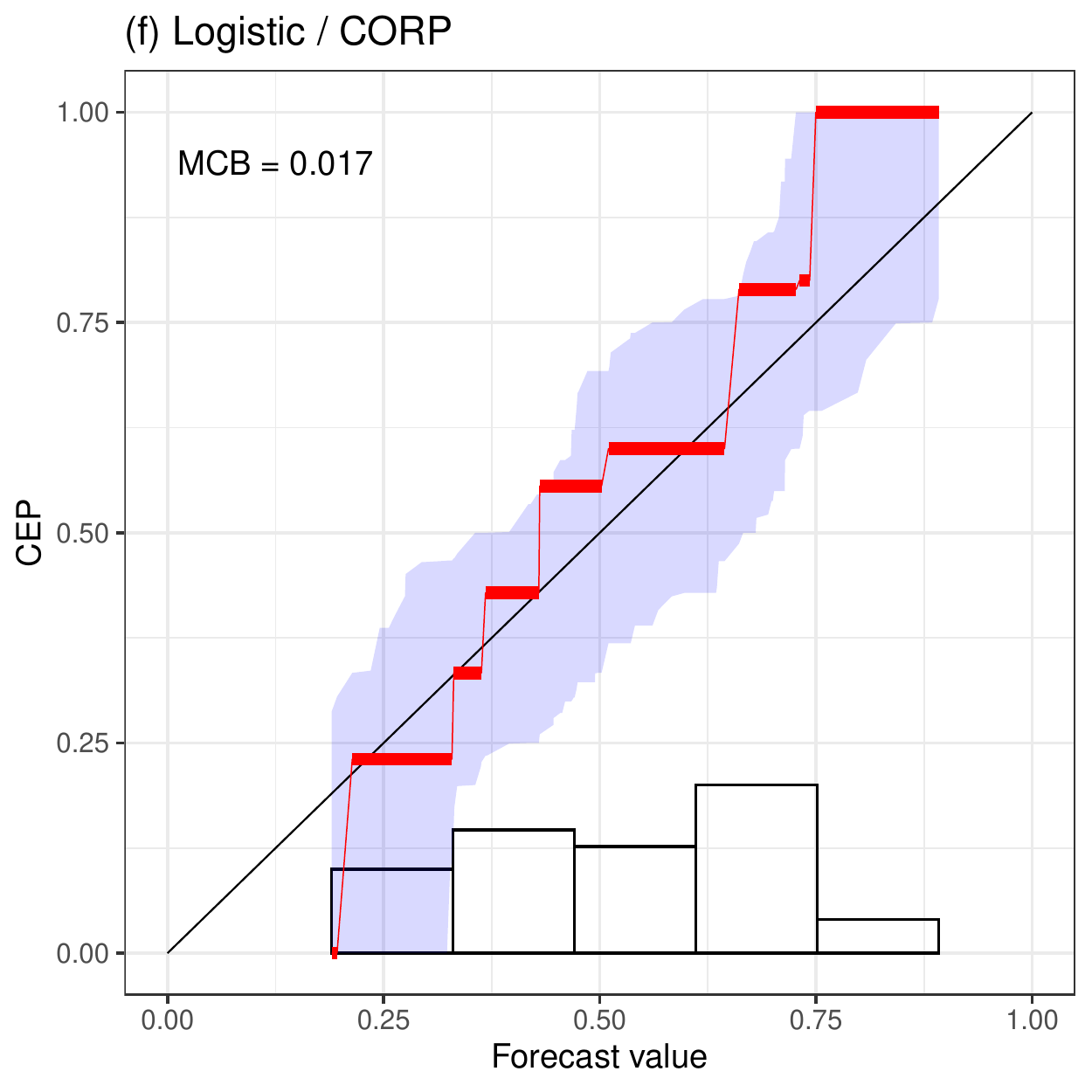}
	\end{subfigure}
	\caption{Reliability diagrams for probability of precipitation
		forecasts over Niamey, Niger (\ref{Vogel2020}) in July--September
		2016 under (a,b) ENS, (c,d) EPC, and (e,f) Logistic methods.  At
		left (a,c,e), we show reliability diagrams under the binning and
		counting approach with a choice of ten equally spaced bins.  At
		right (b,d,f), we show CORP reliability diagrams with uncertainty
		quantification through 90\% consistency bands.  The histograms at
		bottom illustrate the distribution of the $n = 86$ forecast
		values.  \label{fig:Vogel}}
\end{figure}

In Fig.~\ref{fig:Vogel}(a,c,e) we show reliability diagrams based on
the binning and counting approach with a choice of $m = 10$ equally
spaced bins for 24-hour ahead daily probability of precipitation
forecasts at Niamey, Niger in July--September 2016.  They concern
three competing forecasting methods, including the world-leading,
52-member ensemble system run by the European Centre for Medium-Range
Weather Forecasts (ENS, \ref{ECMWF}), a reference forecast called
extended probabilistic climatology (EPC), and a purely data-driven
statistical forecast (Logistic), as described by Vogel et
al.~(\ref{Vogel2020}, Fig.~2).

\begin{figure}[tbh]
	\centering
	\begin{subfigure}{.4\linewidth}
		\includegraphics[width=\linewidth]{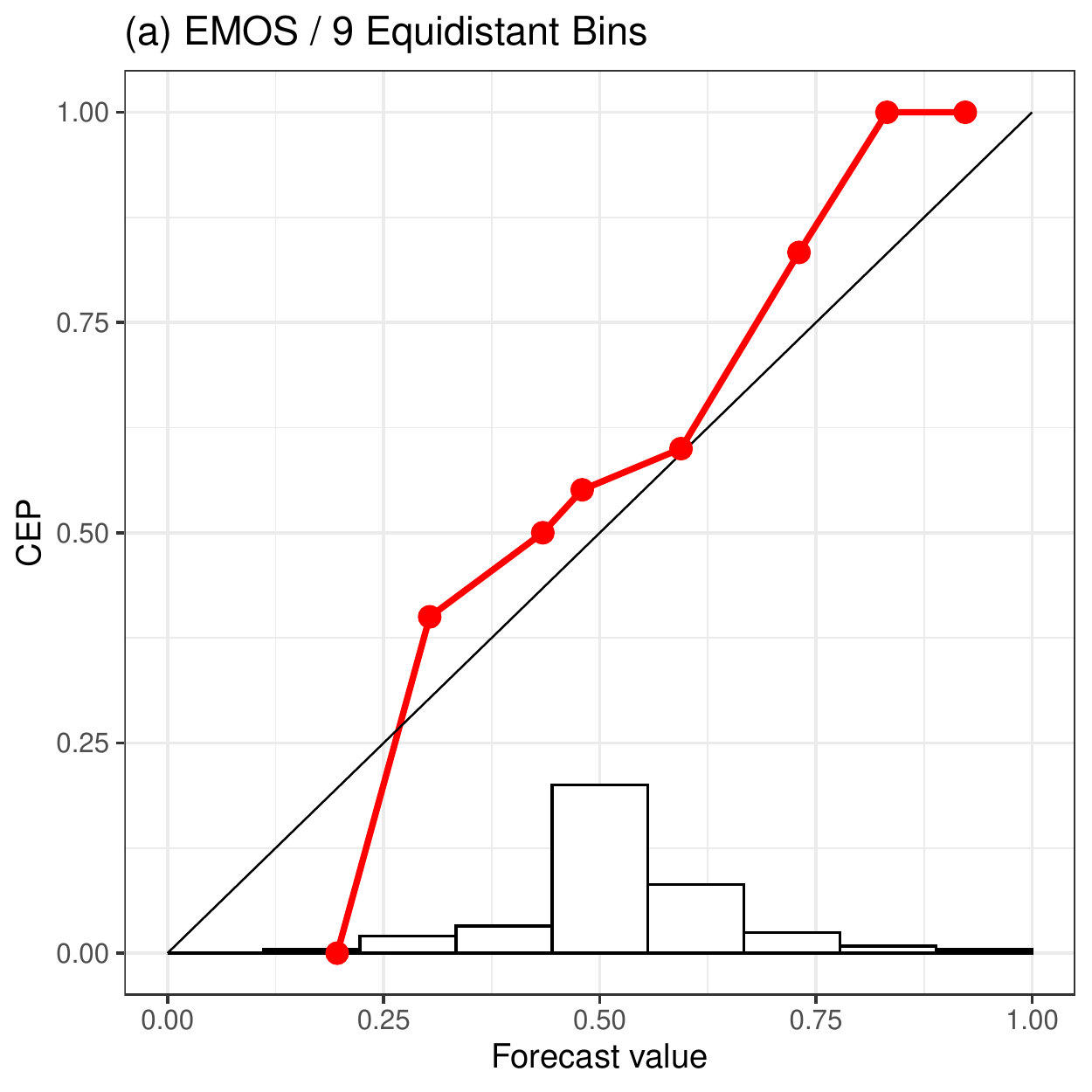} 
	\end{subfigure}
	\begin{subfigure}{.4\linewidth}
		\includegraphics[width=\linewidth]{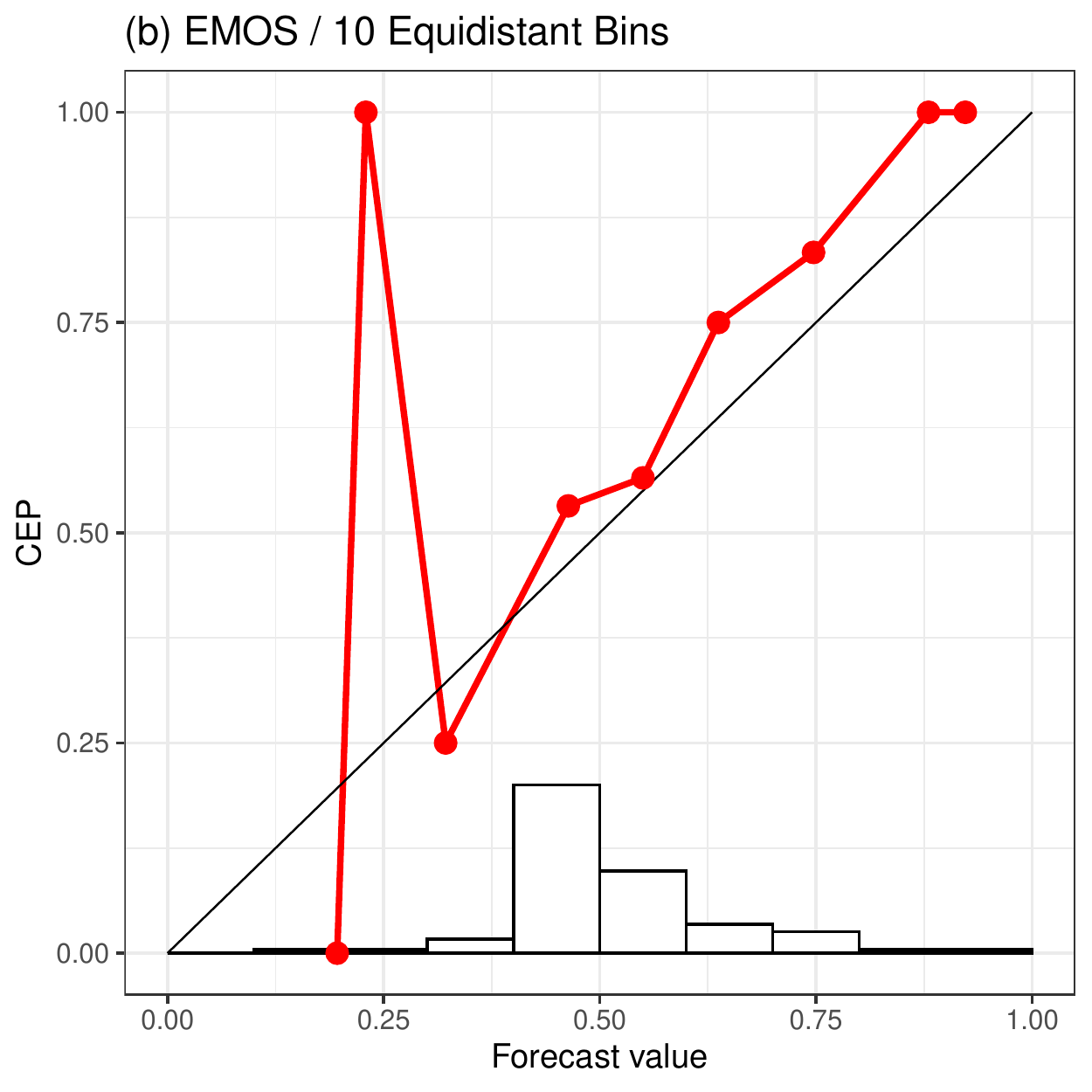} 
	\end{subfigure}
	\begin{subfigure}{.4\linewidth}
		\includegraphics[width=\linewidth]{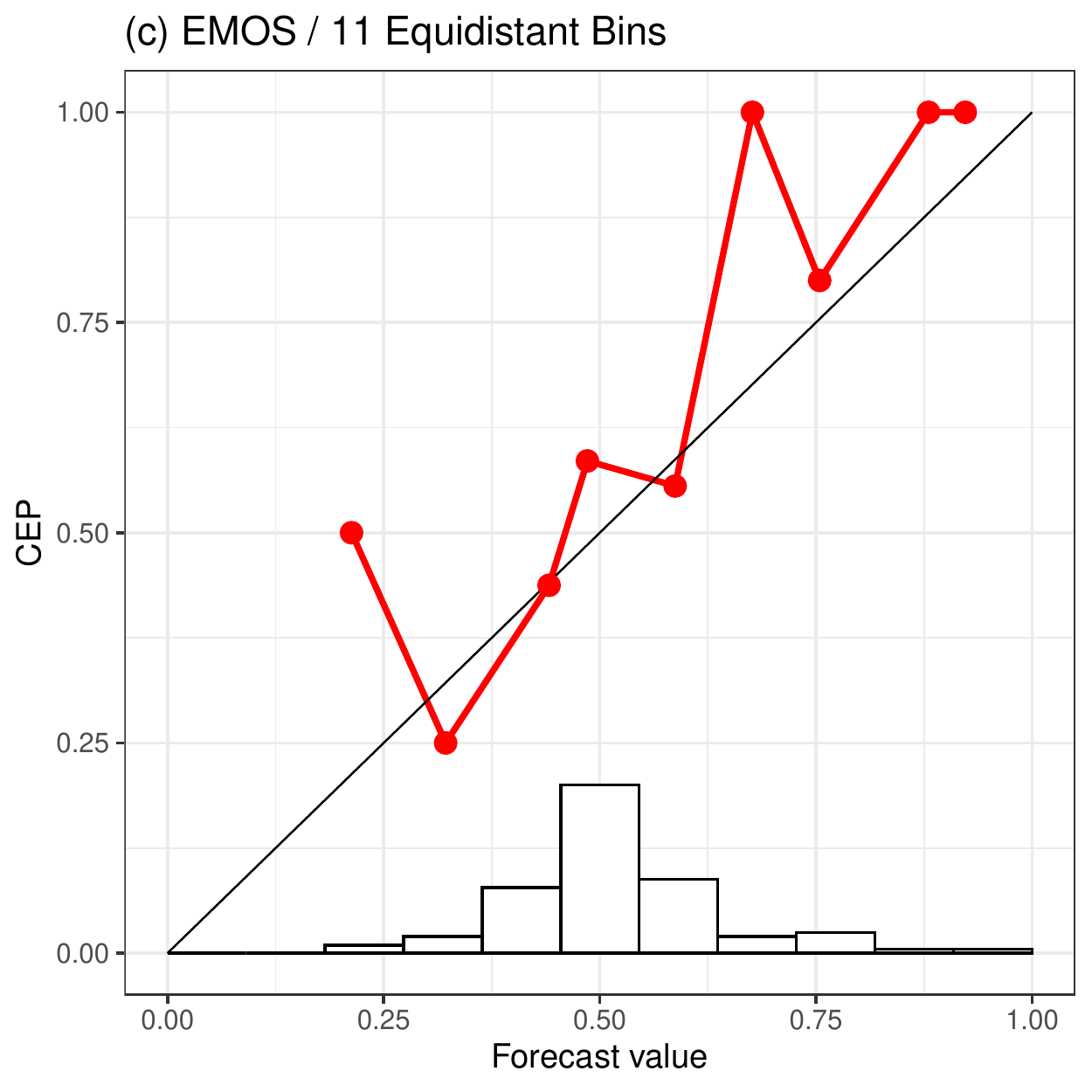} 
	\end{subfigure}
	\begin{subfigure}{.4\linewidth}
		\includegraphics[width=\linewidth]{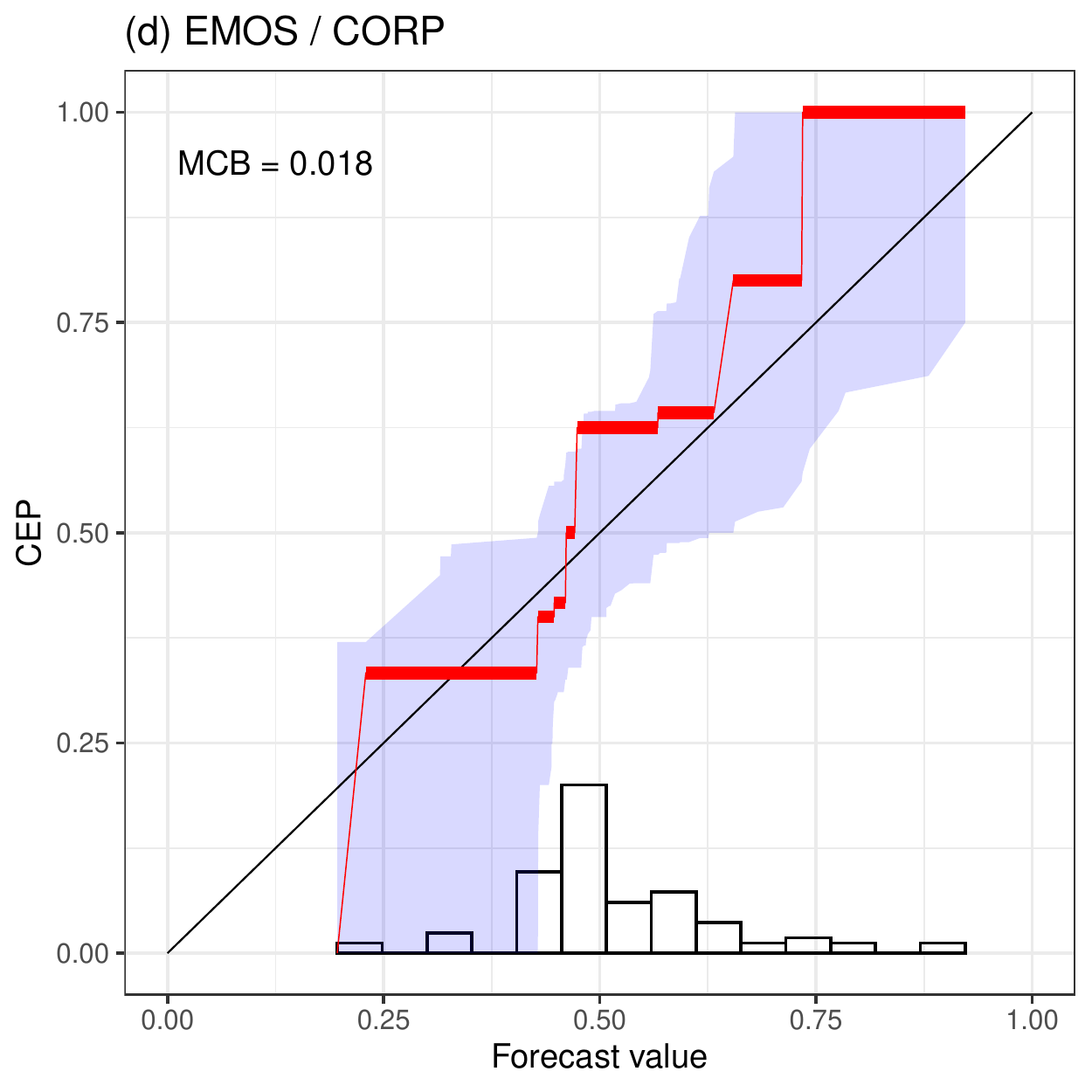} 
	\end{subfigure}
	\caption{Reliability diagrams for probability of precipitation
		forecasts over Niamey, Niger (\ref{Vogel2020}) in July--September
		2016 with the EMOS method, using the binning and counting approach
		with a choice of (a) 9, (b) 10, and (c) 11 equidistant bins,
		together with (d) the CORP reliability diagram, for which we provide
		uncertainty quantification through 90\% consistency
		bands.  \label{fig:stability}}
\end{figure}

Not surprisingly, the classical approach to plotting reliability
diagrams is highly sensitive to the specification of the bins, and the
visual appearance may change drastically under the slightest change.
We show an example in Fig.~\ref{fig:stability}(a--c) for a fourth type
of forecast at Niamey, namely, a statistically postprocessed version
of the ENS forecast called ensemble model output statistics (EMOS),
for which choices of $m = 9$, $10$, or $11$ equidistant bins yield
drastically distinct reliability diagrams.  This is a disconcerting
state of affairs for a widely used data analytic tool, and contrary to
well-argued recent pleas for reproducibility (\ref{Stodden2016}) and
stability (\ref{Yu2020}).  Similar instabilities under the binning and
counting approach have been reported for numerical measures of
calibration, even when the size $n$ of the dataset considered is large
(\ref{Allison2014}, p.~6, \ref{Kumar2019}, Sect.~3.1).  

While methods for the choice of the binning and related implementation
decisions for reliability diagrams have been proposed in the
literature (\ref{Broecker2008}, \ref{Copas1983}, \ref{Atger2004}),
extant approaches lack theoretical justification, are elaborate, and
have not been adopted by practitioners.  Instead, researchers across
discplines continue to craft reliability diagrams and report
associated measures of (mis)calibration, such as the Brier score
reliability component (\ref{Brier1950}--\ref{Kull2015}), based on ad
hoc choices.  In this light, Stephenson et al.~(\ref{Stephenson2008},
p.~757) call for the development of ``nonparametric approaches for
estimating the reliability curves (and hence the Brier score
components), which also include[d] point-wise confidence intervals.''

Here we introduce a new approach to reliability diagrams and score
decompositions, which resolves these issues in a theoretically optimal
and readily implementable way, as illustrated on the forecasts at
Niamey in Figs.~\ref{fig:Vogel}(b,d,f) and \ref{fig:stability}(d).  In
a nutshell, we use nonparametric isotonic regression and the
pool-adjacent-violators (PAV) algorithm to estimate conditional event
probabilities (CEPs), which yields a fully automated choice of bins that
adapts to both discrete and continuous settings, without any need for
tuning parameters or implementation decisions.  We call this stable,
new approach CORP, as its novelty and power include the following four
properties.

\paragraph{Consistency} The CORP reliability diagram and associated numerical 
measures of (mis)\-cal\-i\-bra\-tion are consistent in the classical
statistical sense of convergence to population characteristics.  We
leverage existing asymptotic theory
(\ref{ElBarmi2005}--\ref{Moesching2020}) to demonstrate that the
rate of convergence is best possible, and to generate large sample
consistency and confidence bands for uncertainty quantification.

\paragraph{Optimality} The CORP reliability diagram is optimally binned, 
in that no other choice of bins generates more skillful
(re)calibrated forecasts, subject to regularization via isotonicity
(\ref{BBBB}, Thm.~1.10, \ref{Fawcett2007}, \ref{Bruemmer2013}).

\paragraph{Reproducibility} The CORP approach does not require any tuning 
parameters nor implementation decision, thus yielding well defined
and readily reproducible reliability diagrams and score
decompositions.

\paragraph{Pool-adjacent-violators (PAV) algorithm based} CORP is based on 
nonparametric isotonic regression and implemented via the PAV
algorithm, a classical iterative procedure with linear complexity
only (\ref{Ayer1955}, \ref{deLeeuw2009}).  Essentially, the CORP
reliability diagram shows the graph of the PAV-(re)calibrated
forecast probabilities.

\paragraph{} In the remainder of the article we provide the details of CORP
reliability diagrams and score decompositions, and we substantiate the
above claims via mathematical analysis and simulation experiments.

\section{The CORP approach: Optimal binning via the pool-adjacent-violators (PAV) algorithm} 

The basic idea of CORP is to use nonparametric isotonic regression to
estimate a forecast's CEPs as a
monotonic, non-decreasing function of the original forecast values.
Fortunately, in this simple setting there is one, and only one, kind
of nonparametric isotonic regression, for which the PAV algorithm
provides a simple algorithmic solution (\ref{Ayer1955},
\ref{deLeeuw2009}).  To each original forecast value, the PAV
algorithm assigns a (re)calibrated probability under the regularizing
constraint of isotonicity, as illustrated in textbooks
(\ref{Flach2012}, Figs.~2.13 and 10.7), and this solution is optimal
under a very broad class of loss functions (\ref{BBBB}, Thm.~1.10).
In particular, the PAV solution constitutes both the nonparametric
isotonic least squares and the nonparametric isotonic maximum
likelihood estimate of the CEPs.

The CORP reliability diagram plots the PAV-calibrated probability
versus the original forecast value, as illustrated on the Niamey data
in Figs.~\ref{fig:Vogel}(b,d,f) and \ref{fig:stability}(d).  The PAV
algorithm assigns calibrated probabilities to the individual unique
forecast values, and we interpolate linearly inbetween, to facilitate
comparison with the diagonal that corresponds to perfect calibration.
If a group of (one or more) forecast values are assigned identical
PAV-calibrated probabilities, the CORP reliability diagram displays a
horizontal segment.  The horizontal sections can be interpreted as
bins, and the respective PAV-calibrated probabilities are simply the
bin-specific empirical event frequencies.  For example, we see from
Fig.~\ref{fig:Vogel}(b) that the PAV algorithm assigns a calibrated
probability of $.125$ to ENS forecast values between
$\frac{9}{52}$ and $\frac{20}{52}$, and a calibrated
probability of $.481$ to ENS values between $\frac{21}{52}$
and $\frac{42}{52}$.  The PAV algorithm guarantees that both the
number and the positions of the horizontal segments (and hence the
bins) in the CORP reliability diagram are determined in a fully
automated, optimal way.

The assumption of nondecreasing CEPs is natural, as decreasing
estimates are counterintuitive, routinely being dismissed as artifacts
by practitioners.  Furthermore, the constraint provides an implicit
regularization, serving to stabilize the estimate and counteract
overfitting, despite the method being entirely nonparametric.  Under
the binning and counting approach, small or sparsely populated bins
are subject to overfitting and large estimation uncertainty, as
exemplified by the sharp upward spike at about $.25$ in
Fig.~\ref{fig:stability}(b).  The assumption of isotonicity in CORP
stabilizes the estimate and avoids artifacts
(Fig.~\ref{fig:stability}d).

In contrast to the binning and counting approach, which has not been
subject to asymptotic analysis, CORP reliability diagrams are provably
statistically consistent: If the predictive probabilities and event
realizations are samples from a fixed, joint distribution, then the
graph of the diagram converges to the respective population
equivalent, as a direct consequence of existing large sample theory
for nonparametric isotonic regression estimates
(\ref{ElBarmi2005}--\ref{Moesching2020}).  Furthermore, CORP is
asymptotically efficient, in the sense that its automated choice of
binning results in an estimate that is as accurate as possible in the
large sample limit.  In Appendix B we formalize these arguments and
report on a simulation study, for which we give details in Appendix A,
and which demonstrates that the efficiency of the CORP approach also
holds in small samples.

\begin{figure}[tb]
	\centering
	\begin{subfigure}{.4\linewidth}
		\includegraphics[width=\linewidth]{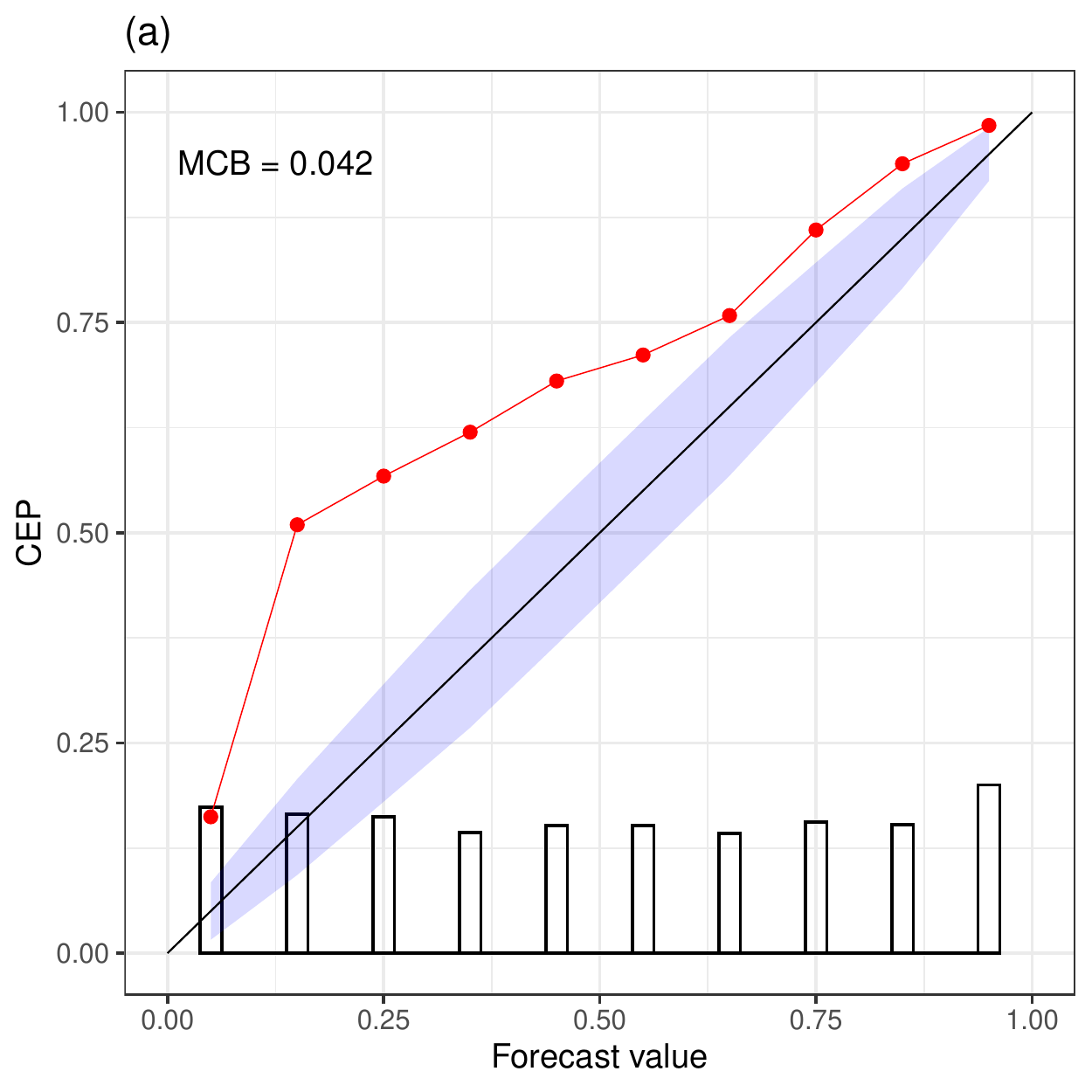} 
	\end{subfigure}
	\begin{subfigure}{.4\linewidth}
		\includegraphics[width=\linewidth]{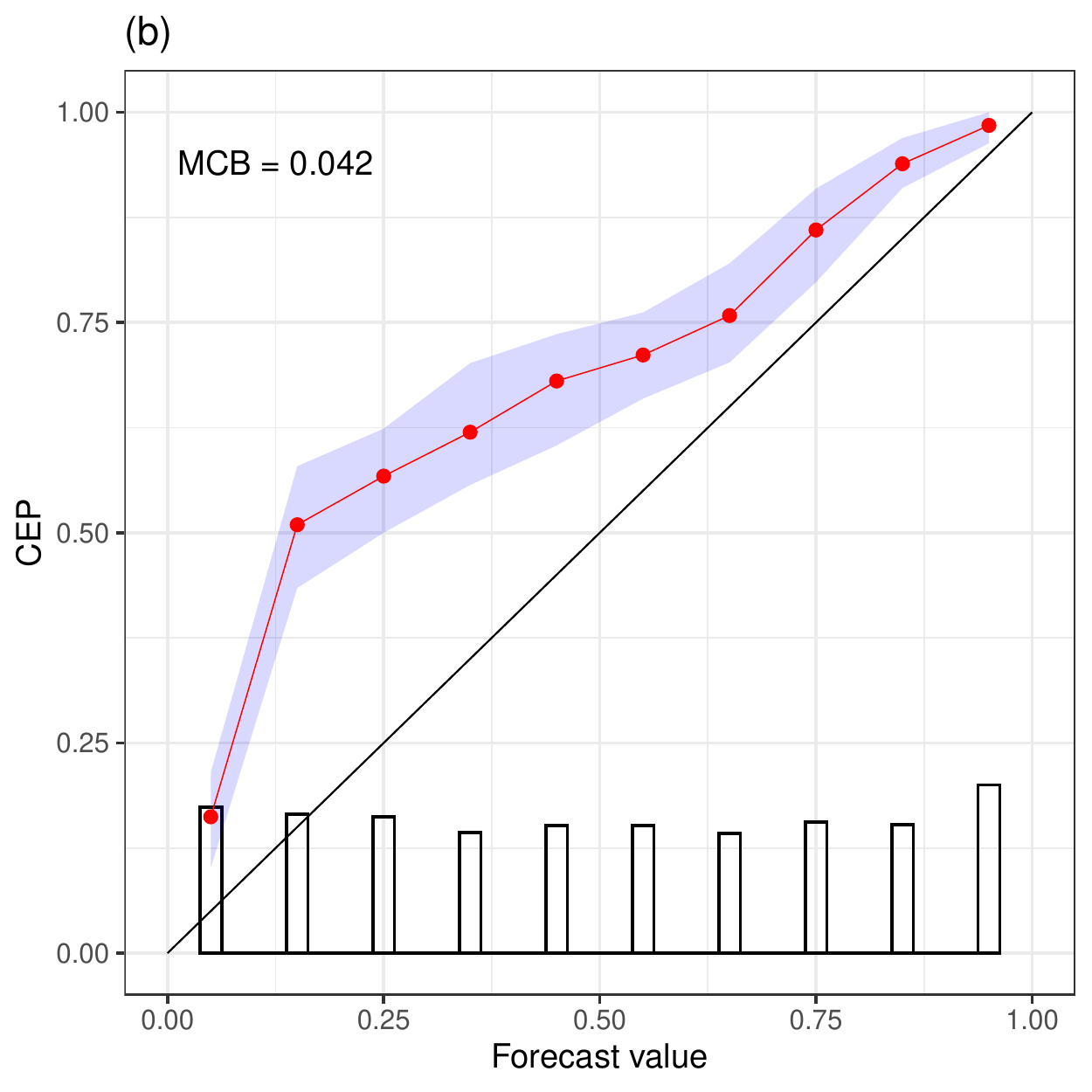} 
	\end{subfigure}
	\begin{subfigure}{.4\linewidth}
		\includegraphics[width=\linewidth]{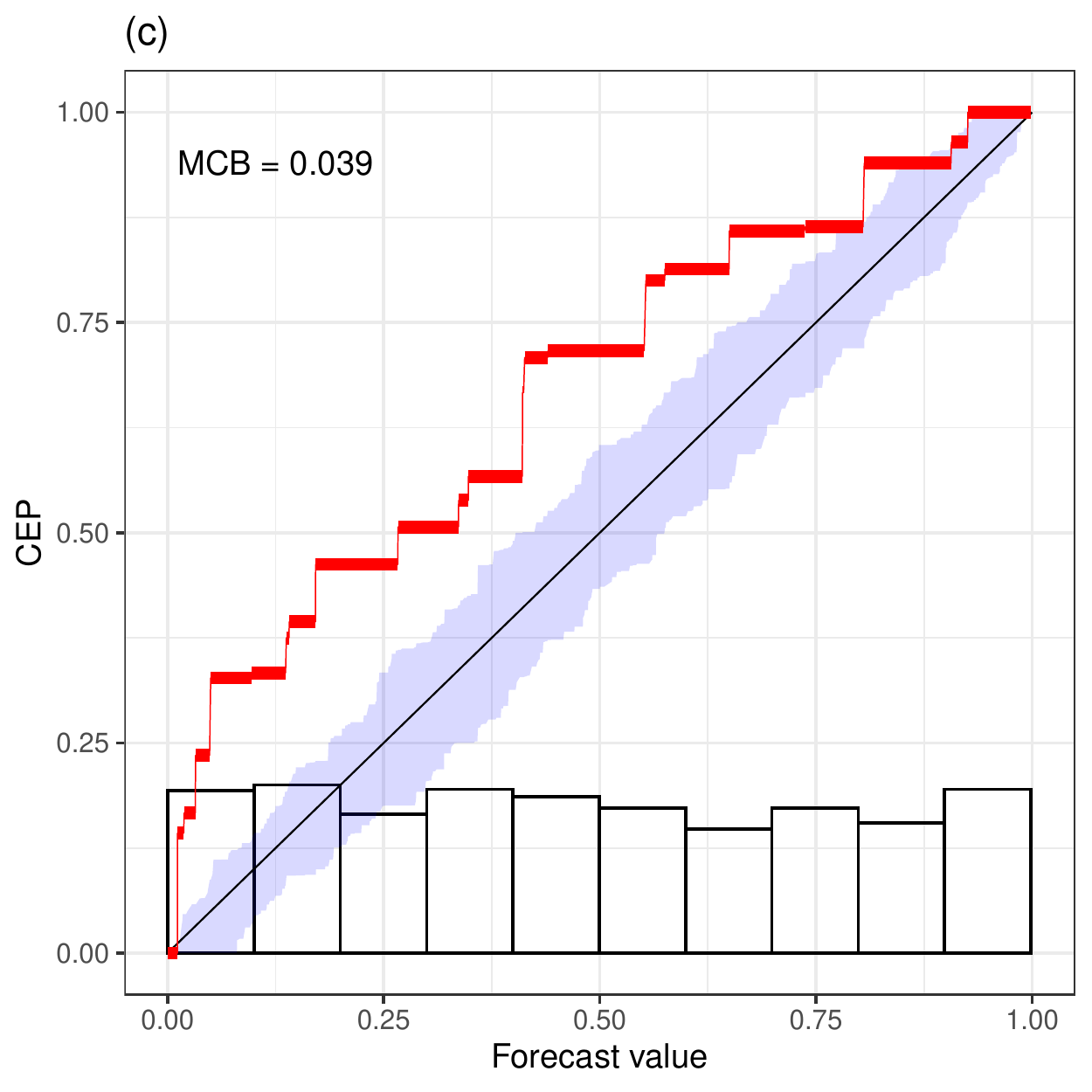} 
	\end{subfigure}
	\begin{subfigure}{.4\linewidth}
		\includegraphics[width=\linewidth]{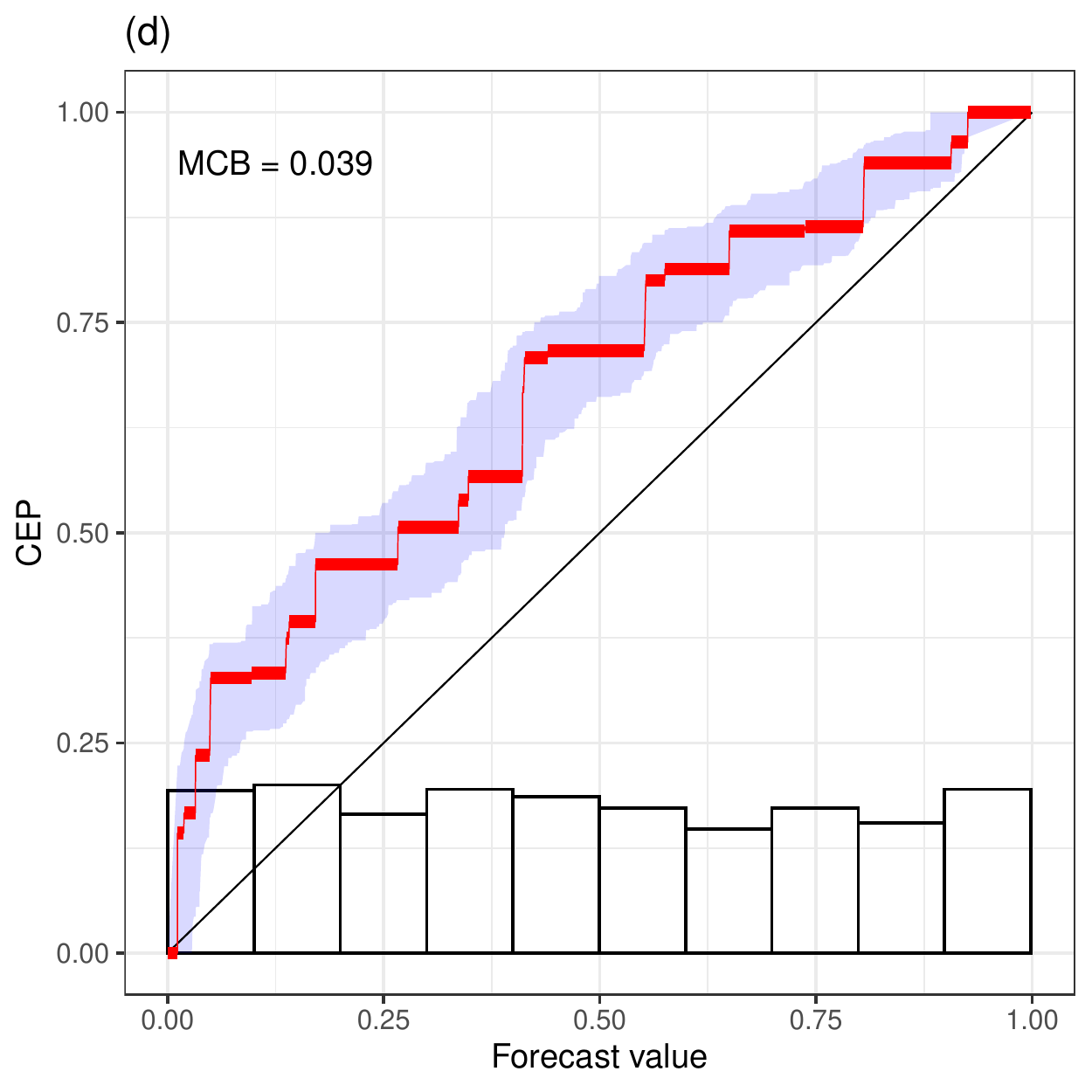} 
	\end{subfigure}
	\caption{CORP reliability diagrams in the setting of (a,b) discretely
		and (c,d) continuously, uniformly distributed, simulated predictive
		probabilities $x$ with a true, miscalibrated CEP of $\sqrt{x}$, with
		uncertainty quantification via (a,c) consistency and (b,d)
		confidence bands at the 90\% level.  \label{fig:main}}
\end{figure}

Traditionally, reliability diagrams have been accompanied by histograms
or bar plots of the marginal distribution of the predictive
probabilities, on either standard or logarithmic scales (e.g.,
\ref{Hamill2008}).  Under the binning and counting approach, the
histogram bins are typically the same as the reliability bins.  In
plotting CORP reliability diagrams, we distinguish discretely and
continuously distributed classifiers or forecasts.  Intuitively, the
discrete case refers to forecast values that only take on a finite and
sufficiently small number of distinct values.  Then we show the
PAV-calibrated probabilities as dots, interpolate linearly inbetween,
and visualize the marginal distribution of the forecast values in a
bar diagram, as illustrated in Fig.~\ref{fig:main}(a,b).  For
continuously distributed forecasts, essentially every forecast takes
on a different value, whence the choice of binning becomes crucial.
The CORP reliability diagram displays the bin-wise constant
PAV-calibrated probabilities in horizontal segments, which are
linearly interpolated inbetween, and we use the Freedman--Diaconis
rule (\ref{Freedman1981}) to generate a histogram estimate of the
marginal density of the forecast values, as exemplified in
Fig.~\ref{fig:main}(c,d).  In our software implementation (\ref{GitPackage}) a simple
default is used: If the smallest distance between
any two distinct forecast values is $0.01$ or larger, we operate in
the discrete setting, and else in the continuous one.  The CORP
reliability diagrams in Figs.~\ref{fig:Vogel}--\ref{fig:main} also
display a new measure of miscalibration (\MCB), discussed in detail
later on as we introduce the CORP score decomposition.

\section{CORP uncertainty quantification} 

Br\"ocker and Smith (\ref{Broecker2007}) convincingly advocate the
need for uncertainty quantification, so that structural deviations of
the estimated CEP from the diagonal can be distinguished from
deviations that merely reflect noise.  They employ a resampling
technique for the binning and counting method in order to find
consistency bands under the assumption of calibration.  For CORP, we
extend this approach in two crucial ways, by generating either
consistency or confidence bands, and by using either a resampling
technique or asymptotic distribution theory, where we leverage
existing theory for nonparametric isotonic regression estimates
(\ref{ElBarmi2005}--\ref{Moesching2020}).

Consistency bands are generated under the assumption that the
probability forecasts are calibrated, and so they are positioned
around the diagonal.  There is a close relation to the classical
interpretation of statistical tests and $p$-values: Under the
hypothesized perfect calibration, how much do reliability diagrams
vary, and how (un)likely is the outcome at hand?  In contrast,
confidence bands cluster around the CORP estimate and follow the
classical interpretation of frequentist confidence intervals: If one
repeats the experiment numerous times, the fraction of confidence
intervals that contain the true CEP approaches the nominal level.  The
two methods are illustrated in Fig.~\ref{fig:main}, where the right
column (b,d) shows confidence bands, and the left column (a,c) shows
consistency bands, as do the CORP reliability diagrams in
Figs.~\ref{fig:Vogel}(b,d,f) and \ref{fig:stability}(d).

In our adaptation of the resampling approach, for each iteration the
resampled CORP reliability diagram is computed, and confidence or
consistency bands are then specified by using resampling
percentiles, in customary ways.  For consistency bands, the resampling
is based on the assumption of calibrated original forecast values,
whereas PAV-calibrated probabilities are used to generate confidence
bands.  While resampling works well in small to medium samples, the
use of asymptotic theory suits cases where the sample size $n$ of the
dataset is large -- exactly when the computational cost of resampling
based procedures becomes prohibitive.  Existing asymptotic theory is
readily applicable and operates under weak conditions on the marginal
distribution of the forecast values, and (strict) monotonicity and
smoothness of (true) CEPs (\ref{ElBarmi2005}--\ref{Moesching2020}).

The distinction between discretely and continuously distributed
forecasts becomes critical here as the asymptotic theory differs
between these cases.  For discrete forecasts, results of El Barmi and
Mukerjee (\ref{ElBarmi2005}) imply that the difference between the
estimated and the true CEP, scaled by $n^{1/2}$, converges to a
(mixture of) normal distribution(s).  For continuous forecasts,
following Wright (\ref{Wright1981}), the difference between the
estimated and the true CEP, magnified by $n^{1/3}$, converges to
Chernoff's distribution (\ref{Groeneboom2001}).  The distinct scaling
laws imply that the convergence is faster in the discrete than in the
continuous case, since in the former the CORP binning stabilizes as it
captures the discrete forecast values, and thereafter the amount of
samples per bin increases linearly, in accordance with the standard
$n^{1/2}$ rate.  In either setting, asymptotic consistency and
confidence bands can be obtained from quantiles of the asymptotic
distributions in customary ways.  As a caveat, both resampling and
asymptotic techniques operate under the assumption of independent, or
at least exchangeable, forecast cases, which may or may not be
warranted in practice.  We encourage follow-up work in dependent data
settings, as recently tackled for related types of data science tools
(\ref{Broecker2020}).

\begin{figure}[tb]
	\includegraphics[width=0.985\linewidth]{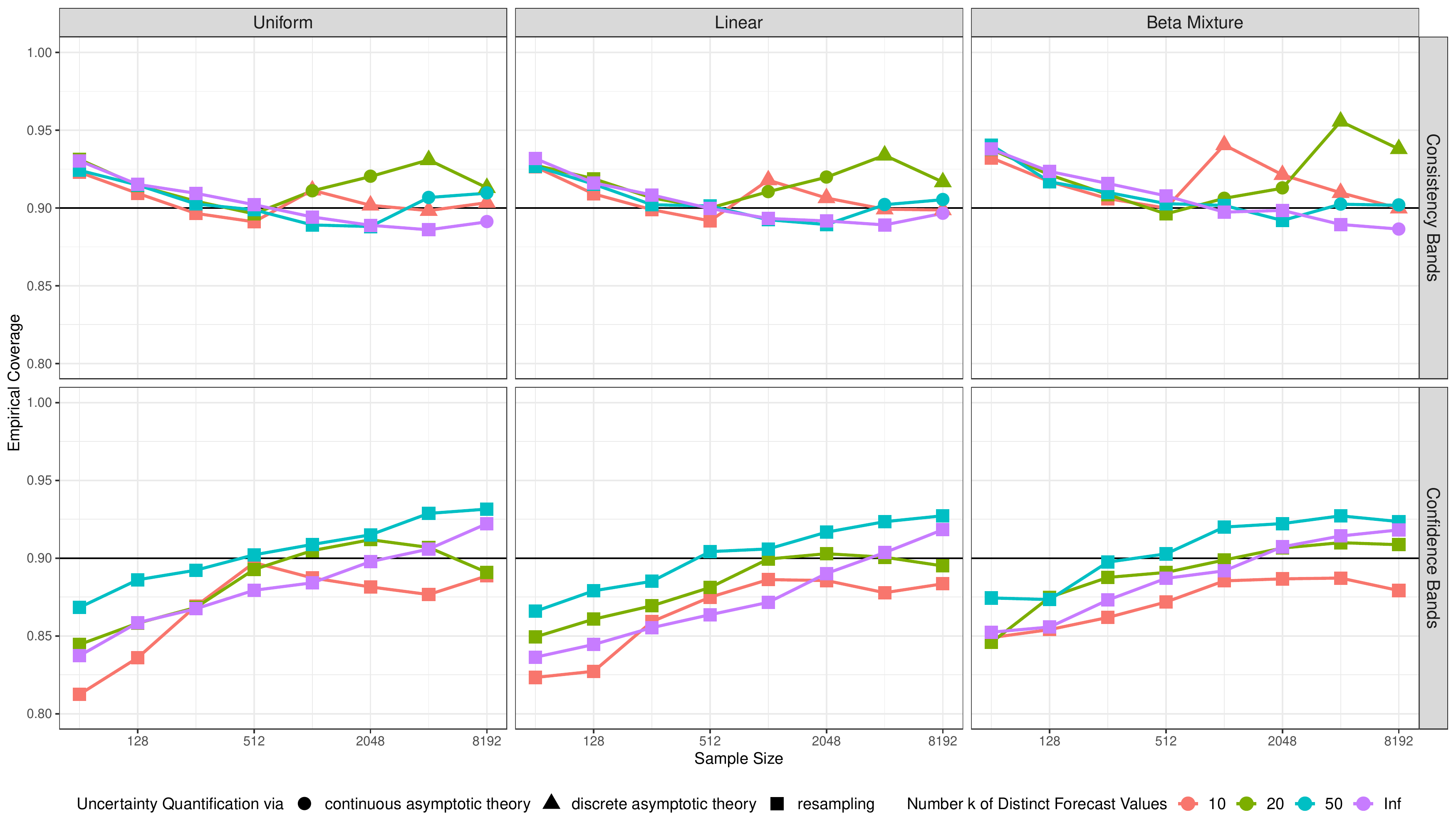}
	\caption{Empirical coverage, averaged equally over the forecast
		values, of 90\% uncertainty bands for CORP reliability diagrams
		under default choices for 1000 simulation replicates.  The upper row
		concerns consistency bands, and the lower row confidence bands.  The
		columns correspond to three types of marginal distributions for the
		forecast values, and colors distinguish discrete and continuous
		settings, as described in Appendix A.  Different symbols denote
		reliance of the bands on resampling, discrete, or continuous
		asymptotic distribution theory. \label{fig:defaults}}
\end{figure}

In our software implementation (\ref{GitPackage}), we use the
following default choices.  Suppose that the sample size is $n$ and
there are $k$ unique forecast values.  For consistency bands, if $n \le
1000$, or if $n \le 5000$ and $n \le 50 k$, we use resampling, else we
rely on asymptotic theory.  In the latter case we employ the discrete
asymptotic distribution if $n \ge 8 k^2$, while otherwise we use the
continuous one.  For confidence bands, the current default uses
resampling throughout, as the asymptotic theory depends on the
assumption of a true CEP with strictly positive derivative.  In the
simulation examples in Fig.~\ref{fig:main}, which are based on $n =
1024$ observations, this implies the use of resampling in panels
(b,c,d) and of discrete asymptotic theory in panel (a).  Fig.~\ref{fig:defaults} 
shows coverage rates of 90\% consistency and
confidence bands in the simulation settings described in Appendix A,
based on the default choices.  The coverage rates are generally
accurate, or slightly conservative, especially in large samples.

\section{CORP score decomposition: Miscalibration (\MCB), discrimination (\DSC), and uncertainty (\UNC) components}

Scoring rules provide a numerical measure of the quality of a
classifier or forecast by assigning a score or penalty $\myS(x,y)$,
based on forecast value $x \in [0,1]$ for a dichotomous event $y \in
\{ 0, 1 \}$.  A scoring rule is proper (\ref{Gneiting2007}) if it
assigns the minimal penalty in expectation when $x$ equals the true
underlying event probability.  If the minimum is unique the scoring
rule is strictly proper.  In practice, for a given sample $(x_1,y_1),
\ldots, (x_n,y_n)$ of forecast-realization pairs the empirical score
\begin{equation}  \label{eq:SX} 
\SX = \frac{1}{n} \sum_{i=1}^n \myS(x_i,y_i)
\end{equation} 
is used for forecast ranking.  Table \ref{tab:scores} presents
examples of proper and strictly proper scoring rules.  The Brier score
and logarithmic score are strictly proper.  In contrast, the
misclassification error is proper, but not strictly proper -- all that
matters is whether or not a classifier probability is on the correct
side of $\frac{1}{2}$.

\begin{table}[t!]
	\footnotesize
	\centering
	\caption{Scoring rules for probabilistic forecasts of binary events \label{tab:scores}}
	\begin{tabular}{l @{\hskip 1cm}  l @{\hskip 1cm}  l}
		\toprule 
		Score                   & Propriety  & Analytic form of $\myS(x,y)$ \\
		\toprule 
		Brier                   & strict     & $(x-y)^2$ \\
		Logarithmic             & strict     & $- y \log x - (1-y) \log(1-x)$ \\
		Misclassification error & non-strict & $\one(x < \frac{1}{2}, y = 1) + \one(x > \frac{1}{2}, y = 0) 
		+ \frac{1}{2} \, \one(x = \frac{1}{2})$ \\
		\bottomrule  
	\end{tabular} 
\end{table}

Under any proper scoring rule, the mean score $\SX$ constitutes a
measure of overall predictive performance.  For several decades,
researchers have been seeking to decompose $\SX$ into intuitively
appealing components, typically thought of as reliability (\REL),
resolution (\RES), and uncertainty (\UNC) terms.  The \REL \ component
measures how much the conditional event frequencies deviate from the
forecast probabilities, while \RES \ quantifies the ability of the
forecasts to discriminate between events and non-events.  Finally, \UNC
\ measures the inherent difficulty of the prediction problem, but does
not depend on the issued forecast under consideration.  While there is
a consensus on the character and intuitive interpretation of the
decomposition terms, their exact form remains subject to debate,
despite a half century quest in the wake of Murphy's
(\ref{Murphy1973}) Brier score decomposition.  In particular, Murphy's
decomposition is exact in the discrete case, but fails to be exact
under continuous forecasts, which has prompted the development of
increasingly complex types of decompositions (\ref{Kull2015},
\ref{Stephenson2008}).

Here we adopt the general score decomposition advocated forcefully by
Siegert (\ref{Siegert2017}), and discussed by various other authors
(e.g., \ref{Kull2015}, \ref{Broecker2009}).  Specifically, let $\SX$, 
\begin{equation}  \label{eq:Sbar} 
\SC = \frac{1}{n} \sum_{i=1}^n \myS(\hat{x}_i,y_i), 
\quad \textrm{and} \quad
\SR = \frac{1}{n} \sum_{i=1}^n \myS(r,y_i)
\end{equation} 
denote the mean score for the original forecast values of
Eq.~[\ref{eq:SX}], the mean score for Calibrated probabilities
$\hat{x}_1, \ldots, \hat{x}_n$, and the mean score for a constant
Reference forecast $r$, respectively.  Then $\SX$ decomposes as
\begin{equation}  \label{eq:decomposition} 
\SX = \underbrace{\left( \SX - \SC \right)}_\MCB 
- \underbrace{\left( \SR - \SC \right)}_\DSC 
+ \underbrace{\SR}_{\UNC},
\end{equation} 
where we adopt, in part, terminology proposed by Ehm and Ovcharov
(\ref{Ehm2017}) and Pohle (\ref{Pohle2020}).  As defined in
Eq.~[\ref{eq:decomposition}], the miscalibration component \MCB \ is
the difference of the mean scores of the original and the calibrated
forecasts.  Similarly, the \DSC \ component quantifies discrimination
ability via the difference between the mean score for the reference
and the calibrated forecast, while the classical measure of
uncertainty (\UNC) is simply the mean score for the reference
forecast.

In the extant literature, it has been assumed implicitly or explicitly
that the calibrated and reference forecasts can be chosen at
researchers' discretion (\ref{Siegert2017}, \ref{Pohle2020}).  We
argue otherwise and contend that the calibrated forecasts ought to be
the PAV-(re)calibrated probabilities, as displayed in the CORP
reliability diagram, whereas the reference forecast $r$ ought to be
the marginal event frequency $\bar{y} = \frac{1}{n} \sum_{i=1}^n y_i$.
We refer to the resulting decomposition as the CORP score
decomposition, which enjoys the following properties:
\begin{itemize} 
	\item $\MCB \geq 0$ with equality if the original
	forecast is calibrated.
	\item $\DSC \geq 0$ with equality if the PAV-calibrated
	forecast is constant.
	\item The decomposition is exact. 
\end{itemize} 
In particular, the CORP score decomposition never yields
counterintuitive negative values of the components, contrary to
choices in the extant literature.  The cases of vanishing components
($\MCB = 0$ or $\DSC = 0$) support the intuitive interpretation of
CORP reliability diagrams, in that parts away from the diagonal
indicate lack of calibration, whereas extended horizontal segments are
indicative of diminished discrimination ability.  For refined
statements and proofs see Theorem 1 in Appendix C.

If $\myS$ is the Brier score, then in the special case of discrete
forecasts with non-decreasing CEPs, the \MCB, \DSC, and \UNC \ terms
in Eq.~[\ref{eq:decomposition}] agree with the \REL, \RES, and \UNC
\ components, respectively, in the classical Murphy decomposition, as
we demonstrate in Theorem 2 in Appendix C.  If $\myS$ is the
misclassification error, \MCB \ equals the fraction of cases in which
the PAV-calibrated probability was on the correct side of
$\frac{1}{2}$, but the original forecast value was not, minus the
fraction vice versa, with natural adaptations in the case of ties.

\begin{table}[t!]
	\centering
	\footnotesize
	\caption{\footnotesize CORP Brier score decomposition for the probability of
		precipitation forecasts in Figs.~\ref{fig:Vogel} and
		\ref{fig:stability}.  \label{tab:Vogel}}
	\begin{tabular}{lcccc}
		\midrule 
		Forecast & $\SX$ & \MCB & \DSC & \UNC \\
		\midrule    
		ENS      & .266  & .066 & .044 & .244 \\
		EPC      & .234  & .022 & .032 & .244 \\
		EMOS     & .232  & .018 & .030 & .244 \\
		Logistic & .206  & .017 & .056 & .244 \\
		\midrule
		\addlinespace
	\end{tabular} 
\end{table} 

In Table \ref{tab:Vogel} we illustrate the CORP Brier score
decomposition for the probability of precipitation forecasts at Niamey
in Figs.~\ref{fig:Vogel}--\ref{fig:stability}.  The purely data-driven
Logistic forecast obtains the best (smallest) mean score, the best
(smallest) \MCB \ term, and the best (highest) \DSC \ component, well
in line with the insights offered by the CORP reliability diagrams,
and attesting to the particular challenges for precipitation forecasts
over northern tropical Africa (\ref{Vogel2020}).

Interestingly, every proper scoring rule admits a representation as a
mixture of elementary scoring rules (e.g., \ref{Gneiting2007},
Sect.~3.2).  Consequently, the \MCB, \DSC, and \UNC \ components of
the CORP decomposition admit analogous representations as mixtures of
the respective components under the elementary scores, whence we may
plot Murphy diagrams in the sense of Ehm et al.~(\ref{Ehm2016}) for
the \MCB, \DSC, and \UNC \ components.  

\section{Discussion} 

Our paper addresses two long-standing challenges in the evaluation of
probabilistic classifiers by developing the CORP reliability diagram
that enjoys theoretical guarantees, avoids artifacts, allows for
uncertainty quantification, and yields a fully automated choice of the
underlying binning, without any need for tuning parameters or
implementation choices.  The associated CORP decomposition
disaggregates the mean score under any proper scoring rule into
components that are guaranteed to be non-negative.

Of particular relevance is the remarkable fact that CORP reliability
diagrams feature optimality properties in both finite sample and large
sample settings.  Asymptotically, the PAV-(re)calibrated
probabilities, which are plotted in a CORP reliability diagram,
minimize estimation error, while in finite samples PAV-calibrated
probabilities are optimal in terms of any proper scoring rule, subject
to the regularizing constraint of isotonicity.

We believe that the proposals in this paper can serve as a blueprint
for the development of novel diagnostic and inference tools for a very
wide range of data science methods.  For example, the popular
Hosmer--Lemeshow goodness-of-fit test (\ref{Hosmer1980}) for logistic
regression is subject to the same types of ad hoc decisions on binning
schemes, and hence the same types of instabilities as the binning and
counting approach (\ref{Allison2014}, p.~6).  Tests based on CORP and
the \MCB \ miscalibration measure are promising candidates for
powerful alternatives.

Perhaps surprisingly, the PAV algorithm and its appealing properties
generalize from probabilistic classifiers to mean, quantile, and
expectile assessments for real-valued outcomes (\ref{Jordan2019}).  In
this light, far-reaching generalizations of the CORP approach apply to
binary regression in general, to standard (mean) regression, where
they yield a new mean squared error (MSE) decomposition with desirable
properties, and to quantile and expectile regression.  In all these
settings, score decompositions have been studied (\ref{Pohle2020},
\ref{Bentzien2014}), and we contend that the PAV algorithm ought to be
used to generate the Calibrated forecast in the general decomposition
in Eq.~[\ref{eq:decomposition}], whereas the Reference forecast ought to be
the respective marginal, unconditional event frequency, mean,
quantile, or expectile.  We leave these extensions to future work and
encourage further investigation from theoretical, methodological, and
applied perspectives.

Open source code for the implementation of the CORP approach in the R
language and environment for statistical computing (\ref{R}) is
available on GitHub (\ref{GitPackage}).

\section*{Appendix A: Simulation settings}

Here we give details for the simulation scenarios in
Figs.~\ref{fig:defaults}--\ref{fig:efficiency}, where we use simple
random samples with forecast values drawn from either Uniform, Linear,
or Beta Mixture distributions, in either the continuous setting, or
discrete settings with $k = 10$, $20$, or $50$ unique forecast values.
The binary outcomes are drawn under the assumption of calibration,
whence the true CEP function coincides with the diagonal.

We begin by describing the continuous setting, where the Uniform
distribution has a uniform density, and the Linear distribution a
linearly increasing density with ordinate $.40$ at $x = 0$ and $1.60$
at $x = 1$.  The Beta Mixture distribution uses Beta$(1,10)$ and
Uniform components with weights $\frac{3}{4}$ and $\frac{1}{4}$,
respectively.  In the discrete settings with $k$ unique forecast
values we maintain the shape of these distributions, but discretize.
Specifically, for $j = 1, \ldots, k$ the probabilistic classifier or
forecast attains the value $x_j = \frac{2j - 1}{2k}$ with probability
\[
p_j = q(x_j) \Big/ \sum_{i=1}^k q(x_i), 
\]
where $q$ is the density in the continuous case.  In
Fig.~\ref{fig:defaults}, we consider discrete settings with $k = 10$,
$20$, and $50$ unique forecast values and the continuous case (marked
Inf).  Fig.~\ref{fig:efficiency} uses discrete settings with $k =
10$ and $50$ unique forecast values and the continuous case.

\section*{Appendix B: Statistical efficiency of CORP}

Suppose that we are given a simple random sample $(x_1,y_1), \ldots,
(x_n,y_n)$ of predictive probabilities $x_1, \ldots, x_n \in [0,1]$
and associated realizations $y_1, \ldots, y_n \in \{ 0, 1 \}$ from an
underlying population, with the true CEP being non-decreasing.

In the case of discretely distributed forecasts that attain a small
number $k$ of distinct values only, results of El Barmi and Mukerjee
(\ref{ElBarmi2005}) imply that the mean squared error (MSE) of the
estimates in a CORP reliability diagram decays at the standard rate of
$n^{-1}$.  If the binning and counting approach separates the distinct
forecast values, the traditional reliability diagram and the CORP
reliability diagram are asymptotically the same, and so are the
respective asymptotic distributions.  However, under the CORP approach
the unique forecast values are always correctly identified as the
sample size increases, while under the binning and counting approach
this may or may not be the case, depending on implementation
decisions.

Large sample theory for the continuously distributed case is more
involved, and generally assumes that the CEP is differentiable with
strictly positive derivative.  Asymptotic results of Wright
(\ref{Wright1981}) for the variance and of Dai et al.~(\ref{Dai2020})
for the bias imply that the MSE of the CORP estimates decays like
$n^{-2/3}$.  We now compare to the binning and counting approach,
using either $m$ fixed, equidistant bins, or using $m = m(n)$
empirical quantile-dependent bins.  For a general sequence of $m(n)$
bins, the magnitudes of the asymptotic variance and squared bias are
governed by the most sparsely populated bin, at a disadvantage
relative to the quantile-dependent case.

The classical reliability diagram relies on a fixed number $m$ of
bins, finds the respective bin-averaged event frequencies, and plots
them against the bin midpoints or bin-averaged forecast values.  Any
such approach fails asymptotically, with estimates that are in general
biased and inconsistent.  More adequately, a flexible number $m(n)$ of
bins can be used, with boundaries defined via empirical quantiles of
$x_1, \dots, x_n$.  Specifically, $m(n)$ bins can be bracketed by 0,
the empirical quantiles at level $j/m(n)$ for $j = 1, \ldots, m(n) -
1$, and 1.  Then, for $n$ sufficiently large, each bin covers about
$n/m(n)$ data points, and the bin-averaged CEPs converge to the true
CEPs at the respective true quantiles with an estimation variance that
decays like $m(n)/n$ and a squared bias that decays like $m(n)^{-2}$.
When $m(n)$ is of order $n^\alpha$ for $\alpha \in (0,1)$, we obtain a
consistent estimate with an estimation variance that decays like
$n^{\alpha - 1}$ and a squared bias that decays like $n^{-2\alpha}$.
Consequently, the MSE of the estimates is of order $n^\beta$ where
$\beta = \max(\alpha - 1, - 2 \alpha)$.  The optimal choice of the
exponent, $\alpha = \frac{1}{3}$, results in an MSE of order
$n^{-2/3}$.  While this asymptotic rate is the same as under the CORP
approach, the CORP reliability diagram is preferable in finite
samples, as we now demonstrate.

\begin{figure}[t]
	\includegraphics[width=0.985\linewidth]{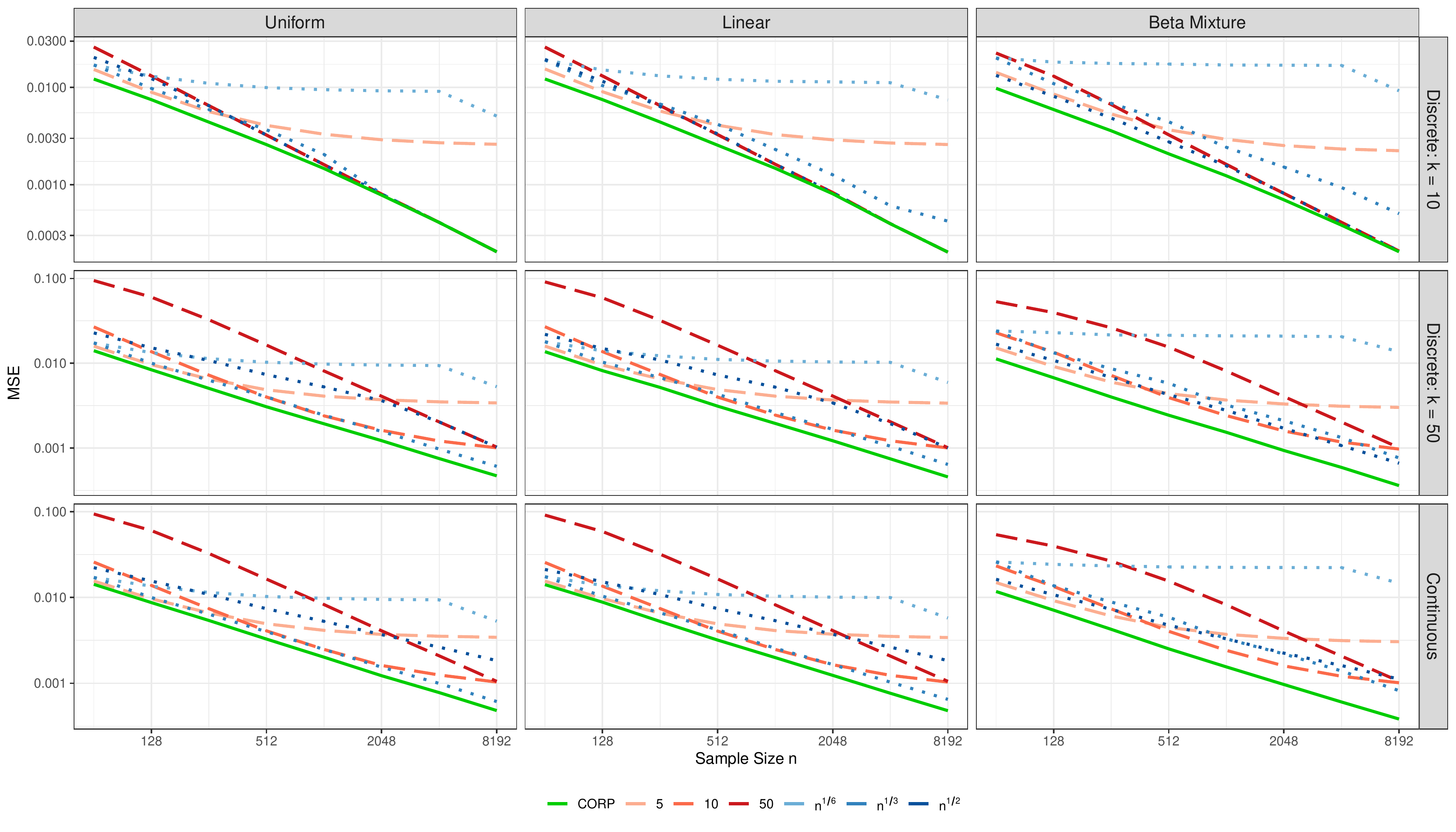}
	\caption{Mean squared error (MSE) of the CEP estimates in CORP
		reliability diagrams for samples of size $n$, in comparison to the
		binning and counting approach with $m = 5$, $10$, or $50$ fixed
		bins, or $m(n) = [n^\alpha]$ quantile-spaced bins, where $\alpha =
		\frac{1}{6}$, $\frac{1}{3}$, or $\frac{1}{2}$.  Note the log-log
		scale.  The simulation settings are described in Appendix A, and MSE
		values are averaged over 1000 replicates. \label{fig:efficiency}}
\end{figure}

In Fig.~\ref{fig:efficiency} we detail a comparison of CORP
reliability diagrams to the binning and counting approach with either
a fixed number $m$ of bins, or $m = m(n) = [n^\alpha]$
empirical-quantile dependent bins, where $[x]$ denotes the smallest
integer less than or equal to $x \in \real$.  For this, we plot the
empirical mean squared error (MSE) of the various CEP estimates
against the sample size $n$, using settings described in Appendix A.
Across colums, the distributions of the forecast values differ in
shape, across rows, we are in the discrete setting with $k = 10$ and
$50$ unique forecasts values, and in the continuous setting,
respectively.  Throughout, the CORP reliability diagrams exhibit the
smallest MSE, uniformly over all sample sizes and against all
alternative methods, with the superiority being the most pronounced
under non-uniform forecast distributions with many unique forecast
values, as frequently generated by statistical or machine learning
techniques.

\section*{Appendix C: Properties of CORP score decomposition} 

Consider data $(x_1,y_1), \ldots, (x_n,y_n)$ in the form of
probability forecasts and binary outcomes, so that $x_1, \ldots, x_n
\in [0,1]$ and $y_1, \ldots, y_n \in \{ 0, 1 \}$.  Let $\bar{y} =
\frac{1}{n} \sum_{i=1}^n y_i$ be the marginal event frequency, and
write $\hat{x}_1, \ldots, \hat{x}_n$ for the PAV-(re)calibrated
probabilities.  Furthermore, let $\SX$, $\SC$, and $\SR$ denote the
mean scores for the original forecast values, (re)calibrated
probabilities, and a reference forecast, as defined in
Eqs.~[\ref{eq:SX}] and [\ref{eq:Sbar}].  With the specific choices of
the PAV-calibrated probabilities as the (re)calibrated forecasts, and
the marginal event frequency $\bar{y}$ as the reference forecast, the
CORP score decomposition in Eq.~[\ref{eq:decomposition}] enjoys the
following properties.

\paragraph{Theorem 1} For every proper scoring rule $\myS$, 
every set of forecast values, and every set of binary outcomes, the
CORP decomposition satisfies the following:
\begin{itemize} 
	\item[(a)] $\MCB = \SX - \SC \geq 0$ with equality if the forecast is calibrated.
	\item[(b)] $\MCB > 0$ if the score is strictly proper and the forecast is uncalibrated.  
	\item[(c)] $\DSC \geq 0$ with equality if the PAV-calibrated forecast is constant. 
	\item[(d)] $\DSC > 0$ if the score is strictly proper and the
	PAV-calibrated forecast is nonconstant.
	\item[(e)] The decomposition is exact. 
\end{itemize} 

\paragraph{Proof} The claims in (a) and (c) rely on the fact that the PAV 
algorithm generates a calibrated forecast that is no worse than the
original forecast in terms of any proper scoring rule (\ref{BBBB},
Thm.~1.10, \ref{Fawcett2007}, \ref{Bruemmer2013}).  If the original
forecast is calibrated, the PAV algorithm leaves it unchanged; if the
PAV algorithm generates a constant forecast, the constant equals the
marginal event frequency $\bar{y}$.

The statements in (b) and (d) follow from the equivalence of (i) and
(iii) in Theorem 2.11 in ref.~(\ref{Gneiting2013}) in
concert with Theorem 3 in ref.~(\ref{Holzmann2014}).
Finally, the claim in (e) is immediate from the definition of the
decomposition. \QEDB

\paragraph{} In the discrete setting we assume that the unique forecasts values 
$z_1 < \cdots < z_k$ are issued $n_1, \ldots, n_k$ times, with $o_1,
\ldots, o_k$ of these cases being events, so that $n_1 + \cdots + n_k
= n$ and $o_1 + \cdots + o_k = n \bar{y}$.  We denote the respective
PAV-calibrated probabilities by $\hat{z}_1 \leq \cdots \leq
\hat{z}_k$.  The classical Brier score decomposition under our choice
of the PAV-calibrated forecast as the calibrated forecast, and
$\bar{y}$ as the reference forecast, then becomes
\[
\SX = \underbrace{\frac{1}{n} \sum_{j=1}^k n_j \left( \frac{o_j}{n_j} - z_j \right)^2}_\REL
- \underbrace{\frac{1}{n} \sum_{j=1}^k n_j \left( \frac{o_j}{n_j} - \bar{y} \right)^2}_\RES
+ \underbrace{\bar{y} \left( 1 - \bar{y} \right)}_\UNC, 
\]
where the \UNC \ component is the same as in the CORP decomposition in
Eq.~[\ref{eq:decomposition}].  Furthermore, subject to mild
conditions, the decompositions agree in full.  

\paragraph{Theorem 2} Under the Brier score, if the 
sequence $o_1/n_1, \ldots, o_k/n_k$ is nondecreasing, then $\MCB =
\REL$ and $\DSC = \RES$, respectively.

\paragraph{Proof} As the sequence $o_1/n_1, \ldots, o_k/n_k$ is nondecreasing, 
the PAV-calibrated probabilities satisfy $\hat{z}_j = o_j/n_j$ for $j
= 1, \ldots, k$.  Adopting the arguments in the Appendix of
ref.~(\ref{Siegert2017}), we see that $\MCB = \SX - \SC = \REL$ and
$\DSC = \RES$. \QEDB

\section*{Data availability} 

The probability of precipitation forecast data at Niamey, Niger are
from the paper by Vogel et al.~(\ref{Vogel2020}, Fig.~2), where the
original data sources are acknowledged.  Reproduction materials,
including data and code in the R software environment (\ref{R}), are
available on GitHub (\ref{GitPackage}, \ref{GitReplMaterial}).

%\acknow{We thank Andreas Fink, Peter Knippertz, Peter Vogel and
%	seminar participants at MMMS2 for providing data, discussion and
%	encouragement.  Our work has been supported by the Klaus Tschira
%	Foundation, by the University of Hohenheim, by the Helmholtz
%	Association, and by the Deutsche Forschungs\-ge\-mein\-schaft (DFG,
%	German Research Foundation) -- Project-ID 257899354 -- TRR 165.}
%
%\showacknow{} % Display the acknowledgments section

\section*{References}

\begin{enumerate}
	\small
	\item \label{Spiegelhalter1986} D.~J.~Spiegelhalter, 
	Probabilistic prediction in patient management and clinical
	trials. \textit{Stat.~Med.}~\textbf{5}, 421--433 (1986).
	\item \label{Murphy1992} A.~H.~Murphy, R.~L.~Winkler, Diagnostic
	verification of probability
	forecasts. \textit{Int.~J.~Forecasting}~\textbf{7}, 435--455 (1992).
	\item \label{Flach2016} P.~A.~Flach, ``Classifier calibration'', in
	\textit{Encyclopedia of Machine Learning and Data Mining} (Springer,
	2016).
	\item \label{Guo2017} C.~Guo, G.~Pleiss, Y.~Sun, K.~Q.~Weinberger,
	``On calibration of modern neutral networks,'' \textit{Proc.~34th
		Int.~Conf. Mach.~Learn.}~(2017).
	\item \label{Broecker2008} J.~Br\"ocker, Some remarks on the
	reliability of categorical probability
	forecasts. \textit{Mon.~Wea.~Rev.}~\textbf{136}, 4488--4502 (2008).
	\item \label{ECMWF} ECMWF Directorate, Describing ECMWF's forecasts
	and forecasting system.  {\em ECMWF Newsl.}~\textbf{133}, 11--13
	(2012).
	\item \label{Vogel2020} P.~Vogel, P.~Knippertz, T.~Gneiting, A.~H.~Fink, 
	M.~Klar, A.~Schlueter, Statistical forecasts for the occurrence of
	precipitation outperform global models over northern tropical
	Africa. Preprint, \url{https://doi.org/10.1002/essoar.10502501.1} (2020).
	\item \label{Stodden2016} V.~Stodden, M.~McNutt, D.~H.~Bailey, E.~Deelman, Y.~Gil, 
	B.~Hanson, M.~A.~Heroux, J.~P.~A.~Ioannidis, M.~Taufer, Enhancing
	reproducibility for computational
	methods. \textit{Science} \textbf{354}, 1240--1241 (2016).
	\item \label{Yu2020} B.~Yu, K.~Kumbier, Veridical data
	science. \textit{Proc.~Natl.~Acad. Sci.~U.S.A.}~\textbf{117},
	3920--3929 (2020).
	\item \label{Allison2014} P.~D.~Allison, Measures of fit for logistic
	regression. Paper 1485-2014, SAS Global Forum, Washington DC (2014).
	\item \label{Kumar2019} A.~Kumar, P.~Liang, T.~Ma, ``Verified
	uncertainty calibration,'' \textit{Proc.~33rd
		Conf.~Neur.~Inform.~Process. Syst.~(NeurIPS)}~(2019).
	\item \label{Copas1983} J.~B.~Copas, Plotting $p$ against $x$. 
	\textit{Appl. Stat.}~\textbf{32}, 25--31 (1983).
	\item \label{Atger2004} F.~Atger, Estimation of the reliability of
	ensemble-based probabilistic
	forecasts. \textit{Q.~J.~R.~Meteorol.~Soc.}~\textbf{130}, 627--646
	(2004).
	\item \label{Brier1950} G.~W.~Brier, Verification of forecasts
	expressed in terms of probability.
	\textit{Mon.~Wea.~Rev.}~\textbf{78}, 1--3 (1950).
	\item \label{Murphy1973} A.~H.~Murphy, A new vector partition of the
	probability score. \textit{J.~Appl.~Meteorol.}~\textbf{12},
	595--600 (1973).
	\item \label{Kull2015} M.~Kull, P.~Flach, ``Novel decompositions of
	proper scoring rules for classification: Score adjustment as
	precursor to calibration'', \textit{ECML PKDD} (2015).
	\item \label{Stephenson2008} D.~B.~Stephenson, C.~A.~S.~Coelho,
	I.~T.~Jolliffe, Two extra components in the Brier score
	decomposition. \textit{Wea.~Forecasting} \textbf{23}, 752--757
	(2008).
	\item \label{ElBarmi2005} H.~El Barmi, H.~Mukerjee, Inferences under a
	stochastic ordering constraint.
	\textit{J.~Am.~Stat.~Assoc.}~\textbf{100}, 252--261 (2005).
	\item \label{Wright1981} F.~T.~Wright, The asymptotic behavior of
	monotone regression estimates. \textit{Ann.~Stat.}~\textbf{9},
	443--448 (1981).
	\item \label{Moesching2020} A.~M\"osching, L.~D\"umbgen, Montone least
	squares and isotonic quantiles. \textit{El.~J.~Stat.}~\textbf{14},
	24--49 (2020).
	\item \label{BBBB} R.~E.~Barlow, D.~J.~Bartholomew, J.~M.~Bremner,
	H.~D.~Brunk, \textit{Statistical Inference under Order Restrictions}
	(Wiley, 1972).
	\item \label{Fawcett2007} T.~Fawcett, A.~Niculescu-Mizil, PAV and the
	ROC convex hull. \textit{Mach.~Learn.}~\textbf{68}, 97--106 (2007).
	\item \label{Bruemmer2013} N.~Br\"ummer, J.~Du Preez, The PAV algorithm
	optimizes binary proper scoring rules. Preprint,
	\url{arXiv:1304.2331} (2013).
	\item \label{Ayer1955} M.~Ayer, H.~D.~Brunk, G.~M.~Ewing, W.~T.~Reid,
	E.~Silverman, An empirical distribution function for sampling with
	incomplete information. \textit{Ann.~Math.~Stat.}~\textbf{26},
	641--647 (1955).
	\item \label{deLeeuw2009} J.~de Leeuw, K.~Hornik, P.~Mair, Isotone
	optimization in R: Pool-adjacent-violators algorithm (PAVA) and
	active set methods. \textit{J.~Stat.~Softw.}~\textbf{32} (2009).
	\item \label{Flach2012} P.~Flach. \textit{Machine Learning: The Art
		and Science of Algorithms that Make Sense of Data} (Cambridge
	University Press, 2012).
	\item \label{Hamill2008} T.~H.~Hamill, R.~Hagedorn, J.~S.~Whitaker, 
	Probabilistic forecast calibration using ECMWF and GFS ensemble
	reforecasts. \textit{Mon.~Wea.~Rev.}~\textbf{136}, 2620--2632
	(2008).
	\item \label{Freedman1981} D.~Freedman, P.~Diaconis, On the histogram
	as a density estimator: $L_2$
	theory. \textit{Z.~Wahr\-schein\-lich\-keitsth.~Verw.~Geb.}~\textbf{57},
	453--476 (1981).
	\item \label{GitPackage}  T.~Dimitriadis, A.~I.~Jordan, R Package 'reliabilitydiag', available at \url{https://github.com/aijordan/reliabilitydiag} (2020).
	\item \label{Broecker2007} J. Br\"ocker, L.~A.~Smith, Increasing the
	reliability of reliability diagrams. \textit{Wea.~Forecasting}
	\textbf{22}, 651--661 (2007).
	\item \label{Groeneboom2001} P.~Groeneboom, J.~A.~Wellner, Computing
	Chernoff's
	distribution. \textit{J.~Computat.~Graph.~Stat.}~\textbf{10},
	388--400 (2001).
	\item \label{Broecker2020} J.~Br\"ocker, Z.~Ben Bouall\`egue, Stratified 
	rank histograms for ensemble forecast verification under serial
	dependence.  \textit{Q.~J.~R.~Meteorol.~Soc.}~\textbf{146}, in press
	(2020).
	\item \label{Gneiting2007} T.~Gneiting, A.~E.~Raftery, Strictly proper
	scoring rules, prediction, and
	estimation. \textit{J.~Am.~Stat.~Assoc.}~\textbf{102}, 359--379 (2007).
	\item \label{Siegert2017} S.~Siegert, Simplifying and generalising
	Murphy's Brier score
	decomposition. \textit{Q.~J.~R.~Meteorol.~Soc.}~\textbf{143},
	1178--1183 (2017).
	\item \label{Broecker2009} J.~Br\"ocker, Reliability, sufficiency, and
	the decomposition of proper
	scores. \textit{Q.~J.~R.~Meteorol.~Soc.}~\textbf{135}, 1512--1519
	(2009).
	\item \label{Ehm2017} W.~Ehm, E.~Y.~Ovcharov, Bias-corrected score
	decomposition for generalized quantiles. \textit{Biometrika}
	\textbf{104}, 473--480 (2017).
	\item \label{Pohle2020} M.-O.~Pohle, The Murphy decomposition and the
	calibration--resolution principle: A new perspective on forecast
	evaluation. Preprint, \url{arXiv:2005.01835} (2020).
	\item \label{Ehm2016} W.~Ehm, T.~Gneiting, A.~Jordan, F.~Kr\"uger, Of
	quantiles and expectiles: Consistent scoring functions, Choquet
	representations and forecast rankings (with
	discussion). \textit{J.~R.~Stat.~Soc.~Ser.~B}~\textbf{78}, 505--562
	(2016).
	\item \label{Hosmer1980} D.~W.~Hosmer, S.~Lemeshow, Goodness-of-fit
	tests for the multiple logistic regression
	Model. \textit{Commun.~Stat.~A}, \textbf{9}, 1043--1069 (1980).
	\item \label{Jordan2019} A.~I.~Jordan, A.~M\"uhlemann, J.~F.~Ziegel,
	Optimal solutions to the isotonic regression problem. Preprint,
	\url{arXiv:1904.04761} (2019).
	\item \label{Bentzien2014} S.~Bentzien, P.~Friederichs, Decomposition
	and graphical portrayal of the quantile
	score. \textit{Q.~J.~R.~Meteorol.~Soc.}~\textbf{140}, 1924--1934
	(2014).
	\item \label{R} R Core Team, R: A language and environment for
	statistical computing. R Foundation for Statistical Computing,
	Vienna, Austria,
	\href{https://www.r-project.org/}{https://www.r-project.org/}
	(2020).
	\item \label{Dai2020} R.~Dai, H.~Song, R.~F.~Barber, G.~Raskutti, The
	bias of isotonic regression.  \textit{El.~J.~Stat.}~\textbf{14},
	801--834 (2020).
	\item \label{Gneiting2013} T.~Gneiting, R.~Ranjan, Combining
	predictive distributions. \textit{El.~J.~Stat.}~\textbf{7},
	1747--1782 (2013).
	\item \label{Holzmann2014} H.~Holzmann, M.~Eulert, The role of the
	information set for forecasting --- with applications to risk
	management. \textit{Ann.~Appl.~Stat.}~\textbf{8}, 595--621 (2014).
	\item \label{GitReplMaterial} T.~Dimitriadis, A.~I.~Jordan, Replication material, available at \url{https://github.com/TimoDimi/replication\_DGJ20} (2020).
\end{enumerate}

\end{document}